\documentclass[journal,twocolumn]{IEEEtran}

\usepackage{color}
\definecolor{gray}{rgb}{0.5,0.5,0.5}

\newcommand{\changed}[1]{\textcolor{blue}{#1}}

\usepackage{amsmath,amsthm}
\usepackage{amssymb,amsfonts}
\usepackage{graphicx}
\usepackage{url}
\usepackage{subfigure}

\newtheorem{theorem}{Theorem}

\newcommand{\argmax}[1]{\mathop{\arg\max}\limits_{#1}}

\newcommand{\mean}[1]{\mathbb{E}\{#1\}}
\newcommand{\meanc}[2]{\mathbb{E}_{#1}\{#2\}}

\newcommand{\tr}[1]{\mathrm{tr}(#1)}

\newcommand{\AkIterd}{\mathbf{A}_{k}^{(d)}}

\newcommand{\DkIterd}{\mathbf{D}_{k}^{(d)}}

\newcommand{\Hk}{\mathbf{h}_{k}}

\newcommand{\Hkmn}{\mathbf{h}_{kmn}}

\newcommand{\IMt}{\mathbf{I}_{M_t}}

\newcommand{\Pone}{\mathbf{p}_{1}}
\newcommand{\Ptwo}{\mathbf{p}_{2}}
\newcommand{\PK}{\mathbf{p}_{K}}
\newcommand{\Pk}{\mathbf{p}_{k}}
\newcommand{\Pl}{\mathbf{p}_{l}}
\newcommand{\PoneIterd}{\mathbf{p}_{1}^{(d)}}
\newcommand{\PtwoIterd}{\mathbf{p}_{2}^{(d)}}
\newcommand{\PKIterd}{\mathbf{p}_{K}^{(d)}}
\newcommand{\PkIterd}{\mathbf{p}_{k}^{(d)}}

\newcommand{\VMt}{\mathbf{V}}

\newcommand{\calRk}{\mathcal{R}_{k}}
\newcommand{\hatHk}{\overline{\mathbf{h}}_{k}}

\newcommand{\sz}{\sigma_z^2}
\newcommand{\sumnok}{\sum\limits_{l \neq k}^K}
\newcommand{\sumK}{\sum\limits_{k =1}^K}

\usepackage{cite}

\ifCLASSINFOpdf
\else
\fi
\begin{document}
%
\title{Robust Linear Precoder Design for 3D Massive MIMO Downlink with A Posteriori Channel Model}

\author{An-An Lu, \IEEEmembership{Member, IEEE,} Xiqi Gao, \IEEEmembership{Fellow, IEEE}
and  Chengshan Xiao, \IEEEmembership{Fellow, IEEE}

\thanks{
Manuscript received November 13, 2021; revised March 19, 2022; accepted March 24, 2022.
The work of A.-A. Lu and X.Q. Gao was supported by  the National Key R\&D Program of China under Grant 2018YFB1801103, the National Natural Science Foundation of China under Grants 61801113, 61960206006 and 61971136, the Jiangsu Province Basic Research Project under Grant BK20195002, the Natural Science Foundation of Jiangsu Province under Grant BK20180362, and the Huawei Cooperation Project.
The work of C. Xiao was supported in part by the US National Science Foundation under Grant ECCS-1827592.
The associate editor coordinating the review of this manuscript
and approving it for publication was Dr. Zhiguo Shi. (\em{Corresponding author: Xiqi Gao.})
}
\thanks{Copyright (c) 2015 IEEE. Personal use of this material is permitted. However, permission to use this material for any other purposes must be obtained from the IEEE by sending a request to pubs-permissions@ieee.org.}
\thanks{A.-A. Lu and X. Q. Gao are with the National Mobile Communications Research Laboratory (NCRL), Southeast University,
Nanjing, 210096 China, and also with Purple Mountain Laboratories, Nanjing 211111, China, e-mail: aalu@seu.edu.cn, xqgao@seu.edu.cn.}
\thanks{C. Xiao is with the Department of Electrical and Computer Engineering, Lehigh University, Bethlehem, PA 18015. Email: xiaoc@lehigh.edu.}
}


\maketitle

\begin{abstract}
In this paper, we investigate the linear precoder design for three dimensional (3D) massive multi-input multi-output (MIMO) downlink with uniform planar array (UPA) and imperfect channel state information (CSI). 
We introduce a beam based statistical channel model (BSCM)  by using sampled steering vectors, and then  
  an {\it{a posteriori}} channel model which includes the channel aging is established.
On the basis of the {\it{a posteriori}} channel model, we consider the robust precoder design by maximizing an upper bound of the expected weighted sum-rate under a total power constraint.
We derive two concave minorizing functions of the objective function. With these minorizing functions and the minorize-maximization (MM) methodology, we derive two iterative algorithms that converge to stationary points of the optimization problem. 
 Simulation results show that
the proposed precoders can achieve a significant performance gain than the widely used regularized zero forcing (RZF) precoder
and the signal to leakage noise ratio (SLNR) precoder in median to high mobility scenarios.

\end{abstract} 

%
\IEEEpeerreviewmaketitle

\section{Introduction}
Massive multiple-input multiple-output (MIMO) \cite{marzetta2016fundamentals,bjornson2014massive}
is one of the enabling
technologies of the fifth generation (5G) mobile networks.
It provides enormous potential capacity gains by employing a large antenna array at a base station (BS),
and enhances multi-user MIMO (MU-MIMO) transmissions on the same time and frequency resource significantly.
With massive antenna arrays at the BS, it is also possible to achieve high energy efficiency.
Furthermore, massive MIMO is a key technology for many new applications and services.
For example, it improves the reliability and the throughput performance for communication with unmanned aerial vehicles (UAVs) \cite{zeng2016wireless}, and well suites for mass connectivity which is very important to support Internet of things (IoT) \cite{liu2018massive}.
There are several types of antenna array in massive MIMO systems.
Among them, the uniform planar array (UPA) is preferred for practical wireless communication systems due to its compact size and three dimensional (3D) coverage ability. In this paper, we investigate the transmission for massive MIMO downlinks with UPA.

To alleviate the multi-user interference and improve the sum-rate performance, the precoders for massive MIMO downlink should be properly designed.
Massive MIMO can be viewed as an extension of conventional multi-user MIMO.
The precoder design for the conventional MU-MIMO and massive MIMO has been widely investigated in different forms over the past years
\cite{peel2005vector,weingarten2006capacity,sadek2007leakage,christensen2008weighted,caire2010multiuser,shi2011iteratively,adhikary2013joint,sunbeam,wang2015asymptotic,park2015multi,liu2016two}.  
The nonlinear precoders such as DPC \cite{weingarten2006capacity} can achieve optimal performance,
but their complexity is very high and thus not suitable to massive MIMO.
Thus, we focus on linear precoder designs for massive MIMO in this paper.
The precoder designs often depend on the available channel state information (CSI) at the BS.
If the BS knows perfect CSI of all user equipments (UEs), the regularized zero forcing (RZF) precoder \cite{peel2005vector} and
the signal to leakage noise ratio (SLNR) precoder \cite{sadek2007leakage} are widely used.
Furthermore, the classic iterative weighted minimum mean square error (WMMSE) precoder \cite{christensen2008weighted,shi2011iteratively}
is designed according to the sum-rate maximization criterion.

In practical massive MIMO systems, perfect CSI at the BS are usually not available due to channel estimation error, channel aging, etc. 
In the literature \cite{caire2010multiuser,park2015multi},
the channel uncertainty are often constructed as a complex Gaussian random matrix with independent and identically distributed (i.i.d.), zero mean and unit variance entries.
However, the uncertainty in practical systems usually deviates from the i.i.d. assumption.
To describe the channel in practical systems more precisely, an \textit{a posteriori} channel model, which  models the time evolution of channel with 
the widely used Gauss-Markov process \cite{mondal2006channel,mamat2018optimizing}, 
is proposed in \cite{lu2017robust}.

In \cite{lu2017robust}, the considered massive MIMO is equipped with a large uniform linear array (ULA).
For such configuration, the jointly correlated channel model \cite{sayeed2002deconstructing,weichselberger2006stochastic,gao2009statistical} with the DFT matrix being the eigenmatrix at the BS side is widely used in the literature. The model is also known as the beam domain channel  model \cite{sayeed2002deconstructing,sunbeam}.
However, for practical massive MIMO with UPA, the number of antennas at each column or each row is usually limited. 
Thus, the accuracy of the straightforward extension of the conventional DFT based channel model is not satisfied.
To overcome this issue, we introduce a beam based statistical channel model (BSCM) by using the matrices of
sampled steering vectors.
Then, we establish an {\it{a posteriori}} beam based statistical channel model which includes the channel aging.

Based on the new  {\it{a posteriori}} channel model, we investigate the robust precoder design for 3D massive MIMO downlink transmission with imperfect CSI. 
To avoid complicated expectation operation involved in the expected weighted sum-rate, we propose to maximize 
its upper bound under a total power constraint.
In this paper, we apply the minorize maximization (MM) methodology  and derive two iterative algorithms for the robust precoder design.
The proposed algorithms converge to the stationary points of the considered optimization problem.

The main contribution of this paper is summarized as follows: 

1. We propose two concave minorizing functions for the upper bound of the expected weighted sum-rate. In the minorizing functions, the precoders of different users are decoupled.  The minorizing functions can be used to apply the MM algorithm, and are easier to optimize in comparison to the original objective function.
 
2. We derive two iterative algorithms for the precoder design based on the two concave minorizing functions and the MM algorithm. The proposed precoders are obtained by solving classic concave optimization problems. Because of the MM methodology, they converge to the stationary points of the considered optimization problem.  

3. The proposed precoders are designed based on the posterior beam based statistical channel models for 3D massive MIMO. Thus, they are  robust to imperfect CSI in 3D massive MIMO systems.
In comparison to the widely used precoders relying on perfect CSI, they can achieve significant performance gain in high mobility scenarios.

The rest of this paper is organized as follows. The channel model is proposed in Section II. The designs of robust linear precoders are presented in
Section III. Simulation results are provided in Section IV. The conclusion is drawn in Section
V.

{\it Notations}: Throughout this paper, uppercase and lowercase boldface letters are used for matrices and vectors, respectively. The superscripts $(\cdot)^*$, $(\cdot)^T$ and $(\cdot)^H$ denote the conjugate, transpose and conjugate transpose operations, respectively. ${\mathbb E}\{\cdot\}$ denotes the mathematical expectation operator.  
The operators ${\rm{tr}}(\cdot)$
 and $\det(\cdot)$ represent the matrix trace and determinant, respectively.  The operator $\otimes$ denotes the Kronecker product. The Hadamard product of two matrices $\mathbf{A}$ and $\mathbf{B}$ is represented by $\mathbf{A} \odot \mathbf{B}$.
The $N \times N$ identity matrix is denoted by $\mathbf{I}_N$.

\section{Channel Model and Problem Formulation}

In this section, we introduce the prior and posterior beam based statistical channel and formulate the considered problem.
\subsection{System Configuration}
\begin{figure} 
\centering
\includegraphics[scale=0.5]{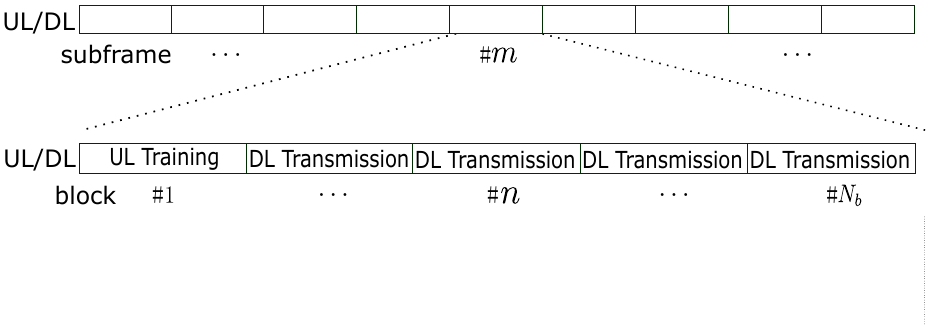}
\caption{Time subframe structure.}
\label{fig:subframe_structure}
\end{figure}

In this subsection, we introduce the system configuration of a 3D massive MIMO system with UPAs equipped in the BS.
We consider a massive MIMO system with block flat fading channels. The massive MIMO system consists of one BS and $K$ UEs. The BS is equipped with an $M_{z} \times M_{x}$  UPA of antennas, where $M_{z}$ and $M_{x}$ are the numbers of pairs of antennas at each vertical column and horizontal row, respectively. Thus, the number of antennas at the BS is $M_t= M_{z}M_{x}$. For simplicity, we assume that all UEs are equipped with single antenna. 
We focus on the case where the massive MIMO system operates in time division duplexing (TDD) mode.
The time subframe structure is plotted in Fig.~\ref{fig:subframe_structure}. We divide the time resources into subframes and each subframe contains $N_b$ blocks.
For simplicity, we omit uplink data transmission, downlink training signal transmission and another signal transmissions, and
assume that there only exists the uplink training phase and the downlink transmission phase. At each subframe, the uplink training sequences are sent once in the first block.  The second block to the $N_b$-th block are used for downlink transmission.
The coherence time is related to the speeds of the users and might be far smaller than the duration of one subframe, and thus  we assume the channel varies from block to block. We also assume the coherence time is larger than one block and the channel coefficients remain constant in a block.

\subsection{Beam Based Statistical Channel Model with Sampled Steering Vectors}
In this subsection, we introduce a beam based  \textit{a priori} statistical channel model.
We denote by $\Hkmn^H$ the channel vector from the BS to the $k$-th UE at the $n$th block of subframe $m$.
\begin{figure}[b]
\centering
\includegraphics[scale=0.4]{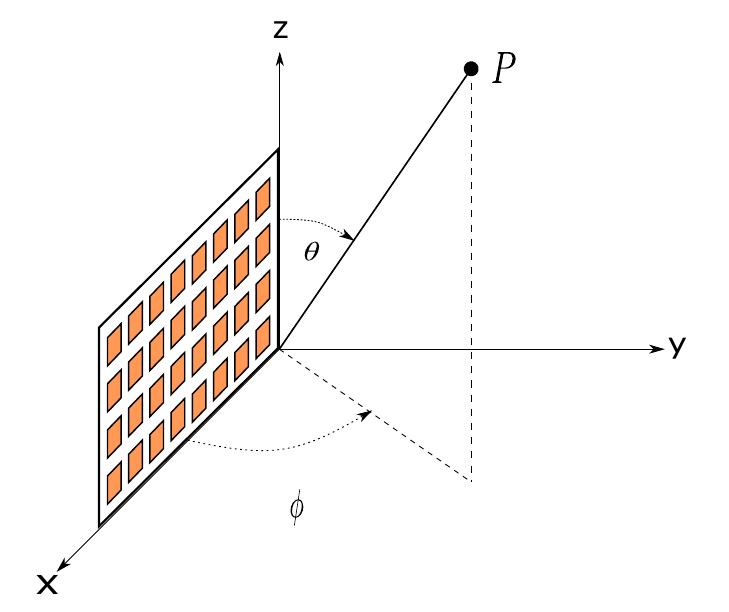}
\caption{3D Coordinate.}
\label{fig:3D corrdinate}
\end{figure}
For the BS, we assume there exists a 3D coordinate as plotted in Fig.~\ref{fig:3D corrdinate}.
The UPA is put on the $xz$-plane.
Let $d_x$ and $d_z$ be the antenna spacing of each row and column of the UPA.
The steering vector $\mathbf{a}_t(u_t,v_t)$ for the UPA at the BS side is extended from that for the ULA as
\begin{equation}
        \mathbf{a}_t(u_t, v_t)=  \mathbf{v}_x(u_t) \otimes \mathbf{v}_z(v_t)
        \label{eq:vector_a_kronecker_product_form}
\end{equation}
where
\begin{IEEEeqnarray}{Cl}
        \mathbf{v}_z(u_t)= \frac{1}{\sqrt{M_z}} [1 ~~ e^{-j2\pi\Delta_z  {u_t}} ~~  \cdots ~~ e^{-(M_{z}-1)j2\pi \Delta_z  {u_t}}]^T \\
        \mathbf{v}_x(v_t)= \frac{1}{\sqrt{M_x}} [1 ~~ e^{-j2\pi \Delta_x {v_t}} ~~  \cdots ~~ e^{-(M_{x}-1)j2\pi \Delta_x  {v_t}}]^T
\end{IEEEeqnarray}
$\Delta_z = \frac{d_z}{\lambda}$, $\Delta_x = \frac{d_x}{\lambda}$, and $u_t$ and $v_t$ are the directional cosines with respect to the $z$ axis and $x$ axis, respectively, \textit{i.e.}, we have  $u_t=\cos\theta_t$ and $v_t=\sin\theta_t\cos\phi_t$. In this paper, we assume that both $d_z$ and $d_x$ equal $0.5\lambda$. Then, we obtain that $\Delta_z= \Delta_x=\frac{1}{2}$.

Similar to that in \cite{tse2005fundamentals} for two dimensional narrowband
MIMO, the multipath channel for the considered 3D massive MIMO from the BS to the $k$-th user at the $n$-th block of the $m$-th subframe can be represented as
\begin{equation}
        \mathbf{h}_{kmn}^H = \sum\limits_{u_t, v_t \in {\cal{B}}_d}g_{kmn}(u_{t}, v_{t})\mathbf{a}_t(u_{t}, v_{t})^H
        \label{eq:orginal_beam_domain_channel_model}
\end{equation}
where  ${\cal{B}}_d$ is the set of directional cosines corresponding to the multiple wireless paths and $g_{kmn}(u_{t}, v_{t})$ is the fading coefficient of each path. 
The channel model provided in \eqref{eq:orginal_beam_domain_channel_model} is the physical channel model,
 where the angle of arrival or departure for each path is arbitrary.
The channel coefficients and the set ${\cal{B}}_d$  are hard to obtain in practice since there are infinitely many possible $u_t, v_t $.
Let ${\cal{B}}$ be the set of all possible directional cosines, and be partitioned into the sets ${\cal{B}}_{t,j,l}$, which is defined as
\begin{IEEEeqnarray}{Cl}
{\cal{B}}_{t,j,l} 
&=\{ (u_t, v_t )|  \|(u_t, v_t)- (u_{t,j}, v_{t,l})\|^2 \le   \nonumber \\
&~~~~~~~~\|(u_t, v_t)- (u_{t,j'}, v_{t,l'})\|^2, \forall j’ \ne j, l’ \ne l\}
\end{IEEEeqnarray}
where $u_{t,j}$, $v_{t,l}$ represent sampled directional cosines, $j=1,2,\cdots, N_z$ and $l=1,2,\cdots,N_x$.

The numbers $N_x$ and $N_z$ are selected as $N_x \geq M_x$ and $N_z \geq M_z$.  Let  $u_{t,j}$ and $v_{t,l}$ be uniformly sampled  in the range $[0~2]$. 
The channel $\mathbf{h}_{kmn}^H$ in $\eqref{eq:orginal_beam_domain_channel_model}$  can be approximated as
\begin{IEEEeqnarray}{Cl}
        \mathbf{h}_{kmn}^H &=\sum\limits_{i,j,l} \sum\limits_{u_t, v_t \in {\cal{B}}_d  \cap  ( {\cal{B}}_{t,j,l})}g_{kmn}(u_{t}, v_{t}) \mathbf{a}_t(u_{t,j}, v_{t,l})^H.
        \label{eq:simple_beam_domain_channel_model}
\end{IEEEeqnarray}

The sampling of the directional cosines and approximating the channel by using the sampled steering vectors have also been used in hybrid analog/digital beamforming (HADB) mmWave massive MIMO systems \cite{lee2016channel, anjinappa2020off}. Here,  an analytical statistical channel model is introduced in the following.
Let $\tilde{g}_{kmn}(u_{t,j}, v_{t,l})$ be defined as
\begin{IEEEeqnarray}{Cl}
   & \tilde{g}_{kmn}(u_{t,j}, v_{t,l})
    =\sum_{u_t, v_t \in {\cal{B}}_d  \cap  (  {\cal{B}}_{t,j,l})}g_{kmn}(u_{t},v_{t}).
\end{IEEEeqnarray}
The channel model in \eqref{eq:simple_beam_domain_channel_model} can be rewritten as
\begin{IEEEeqnarray}{Cl}
    \mathbf{h}_{kmn}^H&= \sum\limits_{j=1}^{N_z}\sum\limits_{l=1}^{N_x }\tilde{g}_{kmn}(u_{t,j}, v_{t,l}) \mathbf{a}_t(u_{t,j}, v_{t,l})^H.
        \label{eq:approximate_beam_domain_channel_model}
\end{IEEEeqnarray}
The fading channels are assumed to be wide-sense stationary uncorrelated scattering (WSSUS) Rayleigh fading, and $\tilde{g}_{kmn}(u_{t,j},v_{t,l})$ are assumed to be independent complex Gaussian random variables with zero means and different variances.
Let  $m_k(u_{t,j},v_{t,l})$ be the positive or negative square root of the variance of $\tilde{g}_{kmn}(u_{t,j},v_{t,l})$ and $w_{kmn}(u_{t,j},v_{t,l})$ be a Gaussian random variable with zero mean and unit variance, then we have that
$\tilde{g}_{kmn}(u_{t,j},v_{t,l})=m_k(u_{t,j},v_{t,l})w_{kmn}(u_{t,j},v_{t,l})$.

In the channel model \eqref{eq:approximate_beam_domain_channel_model}, all users use the same set of sampled steering vectors. 
Furthermore,  each sampled steering vector $\mathbf{a}_t(u_{t,j}, v_{t,l})$ corresponds to a spatial beam.
 Thus, the channel model in \eqref{eq:approximate_beam_domain_channel_model} is called the BSCM, and the channel coefficients $\tilde{g}_{kmn}(u_{t,j}, v_{t,l})$ are called the beam domain channel coefficients.
 Compared with the physical channel model in \eqref{eq:orginal_beam_domain_channel_model}, the channel model 
in \eqref{eq:approximate_beam_domain_channel_model} only includes the sampled steering vectors.

In the following, the model in \eqref{eq:approximate_beam_domain_channel_model} is rewritten as a concrete matrix form.
Let $N_t=N_zN_x$ and $\mathbf{V}$ be the matrix of transmit steering vectors defined by  $\mathbf{V}^H = \mathbf{V}_z^H \otimes \mathbf{V}_x^H \in \mathbb{C}^{N_{t} \times M_{t}}$, where
\begin{IEEEeqnarray}{Cl}
    \mathbf{V}_z&=[\mathbf{v}_z(u_{t,1}), \mathbf{v }_z(u_{t,2}), \cdots, \mathbf{v}_z(u_{t,N_z})] \\
    \mathbf{V}_x&=[\mathbf{v}_x(v_{t,1}), \mathbf{v }_x(v_{t,2}), \cdots, \mathbf{v}_x(v_{t,N_x})].
\end{IEEEeqnarray}
Then, the channel model in \eqref{eq:approximate_beam_domain_channel_model} can be rewritten as
\begin{equation}
        \mathbf{h}_{kmn}^H=\tilde{\mathbf{g}}_{kmn}^H\mathbf{V}^H \label{eq:channel_matrix_correlation_model}
\end{equation}
where $\tilde{\mathbf{g}}_{kmn}^H$ is defined as $[\tilde{\mathbf{g}}_{kmn}^H]_{s}=\tilde{g}_{kmn}( u_{t,j}, v_{t,l})$,
where  $s=jN_x+l$, and is  called the beam domain channel vector. 
For convenience, we also define $[{\mathbf{m}}_{k}]_{is}={m}_{k}(u_{t,j}, v_{t,l})$ and $[{\mathbf{w}}_{kmn}]_{is}={w}_{kmn}(u_{t,j}, v_{t,l})$, where  $s=jN_x+l$.
Then, we have  $\tilde{\mathbf{g}}_{kmn}^H=\mathbf{m}_k^T \odot\mathbf{w}_{kmn}^H$. The vector ${\mathbf{m}}_{k}$ is an $1 \times N_t$ deterministic vector and $\mathbf{w}_{kmn}$ is a complex Gaussian random vector with independent and identically distributed (i.i.d.), zero mean and unit variance entries.
Let $\boldsymbol{\omega}_k$ denote the beam domain channel power matrices as
\begin{equation}
\boldsymbol{\omega}_k={\mathbf{m}}_{k}\odot {\mathbf{m}}_k
\end{equation}
which can be estimated from the samples of the channels   according to
$\mathbb{E}\{[(\mathbf{h}_{kmn}^H\VMt) \odot (\mathbf{h}_{kmn}^H\VMt)^*]\} =\boldsymbol{\omega}_{k}\mathbf{T}_{t}$ where $\mathbf{T}_{t}$ is defined as
\begin{IEEEeqnarray}{Cl} 
     \mathbf{T}_{t}&=\VMt^H\VMt  \odot (\VMt^H\VMt)^* .
     \label{eq:def_of_Ot}
\end{IEEEeqnarray} 
Unlike the conventional DFT based channel model widely used in the literature \cite{sayeed2002deconstructing, you2015pilot,sunbeam}, $\mathbf{V}$   in \eqref{eq:channel_matrix_correlation_model}  is not necessarily a unitary matrix, and the dimension of the beam domain channel vector $\tilde{\mathbf{g}}_{kmn}^H$ is larger than the number of antennas. The matrix $\mathbf{V}$  is   decided by the number of the antennas and the number of the sampled steering vectors. 
When $N_z= M_z$ and  $N_x= M_x$,  the matrix  
$\mathbf{V}$ becomes the Kronecker product of two DFT matrices,  the introduced channel model reduces to the DFT based channel model that straightforwardly extended from ULA  to UPA.
However, its accuracy  is not good enough when  $M_z$ and $M_x$ are of moderate sizes.
To achieve a good complexity-accuracy trade-off, 
the number of sampled steering vectors needs to be selected properly. 
For convenience, we define two fine factors as  $F_z=\frac{N_z}{M_z}$ and  $F_x=\frac{N_x}{M_x}$.

\subsection{A Posteriori Beam based Statistical Channel Model}
The channel model in $\eqref{eq:channel_matrix_correlation_model}$ can be seen as an \textit{a priori} model of the channels before the channel estimation.
The first order Gauss-Markov process  considers the impacts of channel aging \cite{mondal2006channel,mamat2018optimizing}, and can be used to obtain the \textit{a posteriori} CSI of $\tilde{\mathbf{g}}_{kmn}^H$
after the channel estimation.
Since the obtained channel estimation is from the first block, we need to estimate $\tilde{\mathbf{g}}_{km1}^H$.
For simplicity, we assume ideal channel estimation in the first block.
After the channel estimation, we obtain the \textit{a posteriori} CSI of $\tilde{\mathbf{g}}_{kmn}$ as
\begin{equation}
        \tilde{\mathbf{g}}_{kmn}^H= \alpha_{kn}\tilde{\mathbf{g}}_{km1}^H + \sqrt{1-\alpha_{kn}^2}(\mathbf{m}_k^T\odot\mathbf{w}_{kmn}^H)
        \label{eq:posteriori_channel_model}
\end{equation}
where $\alpha_{kn}$ is the temporal correlation coefficient which is related to the moving speed, and $\mathbf{w}_{kmn}$ is a complex Gaussian random matrix with i.i.d., zero mean and unit variance entries.

Finally, the \textit{a posteriori} beam based statistical channel model on the $n$-th block of the $m$-th subframe can be written as
\begin{IEEEeqnarray}{Cl}
        \mathbf{h}_{kmn}^H &= \alpha_{{kn}} \tilde{\mathbf{g}}_{km1}^H \mathbf{V}^H  + \sqrt{1-\alpha_{{kn}}^2} (\mathbf{m}_k ^T \odot \mathbf{w}_{kmn}^H)\mathbf{V}^H \nonumber \\
        &= \overline{\mathbf{h}}_{kmn}^H + \tilde{\mathbf{h}}_{kmn}^H.
        \label{eq:posterori_model_of_Hkm}
\end{IEEEeqnarray}

With \eqref{eq:posterori_model_of_Hkm}, the imperfect CSI for each user is modeled as an {\it a posteriori} statistical channel model with both channel mean and channel variance
information.
The model can describe
the imperfect CSI obtained by the BS in the practical massive MIMO systems under various mobile scenarios.
When $\alpha_{k}$ is very close to 1, the coherence time is large, and the model applies to the quasi-static scenario. When $\alpha_{k}$ becomes very
small, the coherence time is small, and the model can be used for high mobility scenario. By setting the $\alpha_{k}$ according
the mobility of UTs, we are able to describe the channel uncertainties in various  
channel conditions.
Based on the proposed channel model in \eqref{eq:posterori_model_of_Hkm}, we investigate the precoder design robust to the imperfect CSI at the BS in this work.
The proposed posterior channel model  is obtained in the case that the impacts of channel estimation error are not considered. A similar model can be obtained if we  fit the channel estimation error into the  beam based statistical model when the channel estimates are not sufficiently accurate.

In the literature, there also exists another category of research solving the channel aging problem by channel prediction \cite{Kashyap2017Performance,yuan2020machine,Kim2021Massive}, which can obtain  
a similar but more accurate channel model to that in \eqref{eq:posterori_model_of_Hkm}.
The proposed robust precoding method in this work is based on knowing the posterior channel mean and channel covariance matrix. 
It is not contradict with  the channel prediction. On the contrary, we can view the predicted channel as the posterior channel mean, and the prediction error covariance matrix as the posterior channel covariance matrix. Thus, the proposed method developed in this work can be combined with the channel prediction to further improve the sum-rate performance.

\subsection{Problem Formulation}
In this subsection, we present the problem formulation of  the robust linear precoder design. 
We now consider the downlink transmission for the $n$-th block on subframe $m$.
For brevity, we omit the $mn$ in the subscript hereafter.
Let $x_k$ denote the transmitted symbol to the $k$-th UE at the $n$-th block of subframe $m$. The covariance of $x_k$ is one.
The received signal $y_{k}$ at the $k$-th UE for a single symbol interval  can be written as
\begin{equation}
        y_{k} = \mathbf{h}_{k}^H\mathbf{p}_{k}x_{k}+\mathbf{h}^H_{k}\sumnok\mathbf{p}_{l}x_{l} + z_{k}
\end{equation}
where $\mathbf{p}_k$ is the $M_t \times 1$ precoding vector of the $k$-th user, and $z_{k}$ is a
complex Gaussian noise symbol distributed as $\mathcal{CN}(0,\sz )$.

In the literature, an often used method to deal with the imperfect CSI is to treat the uncertainty component $\tilde{\mathbf{h}}_{k}^H\mathbf{p}_{k}x_{k}$ in  $\mathbf{h}_{k}^H\mathbf{p}_{k}x_{k}$ as noise \cite{aquilina2017weighted}.
It works well when the uncertainty part of the channel is relatively small.
However, for the considered channel model, $\hatHk$ can even be close to zero if severe channel aging caused by high mobility happens. 
In such case, treating the uncertainty component $\tilde{\mathbf{h}}_{k}^H\mathbf{p}_{k}x_{k}$ as noise will deteriorate the system performance severely.

On the basis of the {\it a posteriori} channel model, we follow a different way to design the linear precoder, and consider the expected rate as the objective.
We assume that the UEs obtain the perfect CSI of their corresponding channel  $\Hk^H\Pk$ from the precoding domain training signals.  The DL training phase is   omitted in the subframe structure for simplicity. At each UE, we treat the aggregate interference-plus-noise $z'_k =  \Hk^H\sum_{l\ne k}^K\Pl x_l + z_k$ as Gaussian noise.
Let $v_k$ denote the variance of $z'_k$, we have that
    \begin{equation}
        v_k = \sz +  \sumnok\mean{\Hk^H\Pl\Pl^H\Hk}.
        \label{eq:definition_of_covariance_matrices_of_effective_noise}
    \end{equation}
We assume the covariance $v_k$ is known at the $k$-th user.
In such case,
the expected rate of the $k$-th user at subframe $m$ is given by
    \begin{IEEEeqnarray}{Cl}
          \calRk  
           &= \meanc{}{\log({ 1+ v_k^{-1}\Hk^H\Pk\Pk^H\Hk})}
          \label{mutual_information}
    \end{IEEEeqnarray}
where $\meanc{}{\cdot}$ can be computed according to the {\it  a posteriori}  channel model provided in \eqref{eq:posterori_model_of_Hkm}.


Since the $\log\det(\cdot)$ is a convex function, we can obtain
an upper bound of the expected rate of each user as
    \begin{IEEEeqnarray}{Cl}
          \calRk^{ub}  =  \log({ 1 + v_k^{-1}\mean{\Hk^H\Pk\Pk^H\Hk}}).
          \label{mutual_information_approximation}
    \end{IEEEeqnarray}
Compared with the expected rate, the upper bound is easier to compute.
In this work, we are interested in finding the precoding vectors $\Pone,\Ptwo,\cdots,\PK$ that maximize the upper bound of the weighted sum-rate. The optimization problem can be formulated as 
\begin{IEEEeqnarray}{Cl}
        \Pone^{\diamond},\Ptwo^{\diamond},\cdots,\PK^{\diamond}
            =&\argmax{\Pone,\cdots,\PK}\sumK w_k\calRk^{ub}
        \nonumber \\
        &~~~~  {\rm s.t.} ~~  \sumK \tr{\Pk\Pk^H} \leq 1
        \label{eq:objective_optimization_problem_upper_bound}
\end{IEEEeqnarray}
 where $w_k$ is the weight to ensure fairness among the users.
 
\section{Robust Linear Precoder Design}

In this section, we first derive two concave minorizing functions of the objective function. Then, we propose two iterative algorithms for the precoder design based on the minorizing functions.

 \subsection{Concave Minorizing Functions}
 Since the precoders of different users are coupled in the expression of the rate, the upper bound of the expected weighted sum-rate in \eqref{eq:objective_optimization_problem_upper_bound} 
is still difficult to be optimized directly. In the following, we propose two concave minorizing functions for the objective function.

We first provide the definition of a minorizing function, which is usually related to iterative algorithms.
 Let $\PoneIterd,\PtwoIterd, \cdots, \PKIterd$ be the precoding vectors at the $d$-th iteration and let
    \begin{equation}
        g(\Pone,\Ptwo, \cdots, \PK|\PoneIterd,\PtwoIterd, \cdots, \PKIterd) \nonumber
    \end{equation}
denote a real-valued continuous function of the precoders $\Pone,\Ptwo, \cdots, \PK$ whose expression depends on the  precoding vectors $\PoneIterd,\PtwoIterd, \cdots, \PKIterd$ at the $d$-th iteration.
The function $g$ is called a minorizing function of $f$ at the points $\PoneIterd,\PtwoIterd, \cdots, \PKIterd$
provided that\cite{hunter2004tutorial}
\begin{IEEEeqnarray}{Cl}
        &g(\Pone,\Ptwo, \cdots, \PK|\PoneIterd,\PtwoIterd, \cdots, \PKIterd) \nonumber \\
        &~~ \leq  f(\Pone,\Ptwo, \cdots, \PK)
        \label{eq:minorize_condition_1}
\end{IEEEeqnarray}
where the equality holds when we have that $\Pone,\Ptwo, \cdots, \PK=\PoneIterd,\PtwoIterd, \cdots, \PKIterd$.


Let $\mathbf{R}_k$ be defined as $\mathbf{R}_k = \mean{\Hk\Hk^H}$.
From this definition and  \eqref{eq:definition_of_covariance_matrices_of_effective_noise}, we obtain that 
    \begin{IEEEeqnarray}{Cl}
        v_k &= \sz +  \sumnok\mean{\Pl^H\Hk\Hk^H\Pl} \nonumber \\
        &=\sz +  \sumnok \Pl^H\mathbf{R}_k \Pl.
    \end{IEEEeqnarray}
For convenience, we also define $\check{v}_k$
\begin{equation}
	\check{v}_k = v_k+   \mean{\Hk^H\Pk\Pk^H\Hk}.
	 \label{eq:definition_of_checkr}
\end{equation}
Recall that $\hatHk$ denote  the mean of $\mathbf{h}_{k}$, and
$\tilde{\mathbf{h}}_{k}^H$ denote the random part of $\mathbf{h}_{k}$.
Then, $\mathbf{R}_k$  can be computed by
\begin{IEEEeqnarray}{Cl}
    \mathbf{R}_k &= \hatHk\hatHk^H + \meanc{}{\tilde{\mathbf{h}}_{k}\tilde{\mathbf{h}}_{k}^H}  \nonumber \\
    &=  \hatHk\hatHk^H + (1-\alpha_k^2)\VMt{\boldsymbol{\Lambda}}_{k}\VMt^H  
\end{IEEEeqnarray}
where  the second equality is due to $\tilde{\mathbf{h}}_{k}^H=\sqrt{1-\alpha_{\changed{k}}^2} (\mathbf{m}_k ^T \odot \mathbf{w}_{k}^H)\mathbf{V}^H$ from \eqref{eq:posterori_model_of_Hkm}, and ${\boldsymbol{\Lambda}}_{k}$  is a diagonal matrix defined by ${\boldsymbol{\Lambda}}_{k}={\rm diag}(\boldsymbol{\omega}_k)$. 
 Hereafter, we use the notation  $(d)$ in the superscript to denote the object with the condition $\Pone,\Ptwo, \cdots, \PK=\PoneIterd,\PtwoIterd, \cdots, \PKIterd$. For example, $v_k^{(d)}$ is obtained as
    \begin{IEEEeqnarray}{Cl}
        v_k^{(d)}
        &=\sz +  \sumnok (\Pl^{(d)})^H\mathbf{R}_k \Pl^{(d)}.
    \end{IEEEeqnarray}
We then obtain the minorizing function $g_1$ provided in the following theorem.
\begin{theorem}
\label{th:minorizing_function_1}
Let $g_1$ be a function defined as
    \begin{IEEEeqnarray}{Cl}
 g_1 &= c_1^{(d)}
            + \sumK w_k(\mathbf{p}_k^{(d)})^H\mathbf{A}_k^{(d)} \mathbf{p}_k  \nonumber \\
            &~~ +\sumK w_k\mathbf{p}_k^H\mathbf{A}_k^{(d)}\mathbf{p}_k^{(d)}   - \sumK \mathbf{p}_k^H\mathbf{D}_k ^{(d)}\mathbf{p}_k
\end{IEEEeqnarray}
where $c_1^{(d)}$ is a constant and the matrices  $\mathbf{A}_k$ and $\mathbf{D}_k$ is defined as
\begin{IEEEeqnarray}{Cl}
        \mathbf{A}_k  &= v_k^{-1}\mathbf{R}_k.
        \label{eq:definition_of_Ak}
        \\
          \mathbf{B}_k &=    v_k^{-1}\check{v}_k^{-1}\mathbf{R}_k\Pk\Pk^H\mathbf{R}_k
        \label{eq:computation_of_Bk}
             \\
          \mathbf{C}_k &=    (v_k^{-1}-\check{v}_k^{-1})\mathbf{R}_k
           \label{eq:computation_of_Ck} 
           \\
           \mathbf{D}_k &=   w_k\mathbf{B}_k + \sumnok w_l\mathbf{C}_l.
           \label{eq:computation_of_Dk}
\end{IEEEeqnarray}
Then, $g_1$ is a minorizing function of $f$ at the points $\PoneIterd,\PtwoIterd, \cdots, \PKIterd$.
\end{theorem}
\begin{IEEEproof}
The proof is provided in Appendix \ref{sec:proof_of_minorizing_function_1}. 
\end{IEEEproof}
The minorizing function $g_1$ provided in 
Theorem \ref{th:minorizing_function_1} is a concave function of the precoders. Furthermore, the precoders of different users are decoupled in this function. Thus, it is much easier to optimize than the original objective function.
In $g_1$, the matrix $\mathbf{D}_k$ is different for different user, which will cause high computational complexity when developing algorithms.
To make the optimization problem easier to solve, we present an alternative minorizing function modified from the minorizing function $g_1$ in the following theorem.

\begin{theorem}
\label{th:minorizing_function_2}
Let $g_2$ be a function defined as
    \begin{IEEEeqnarray}{Cl}
 g_2 &= c_2^{(d)}
            + \sumK w_k(\mathbf{p}_k^{(d)})^H\mathbf{A}_k^{(d)} \mathbf{p}_k  +\sumK w_k\mathbf{p}_k^H\mathbf{A}_k^{(d)}\mathbf{p}_k^{(d)}  \nonumber \\
            & ~~~~- \sumK \mathbf{p}_k^H\mathbf{D}^{(d)}\mathbf{p}_k
\end{IEEEeqnarray}
where $c_2^{(d)}$ is a constant and the matrix $\mathbf{D}$ is defined as
\begin{IEEEeqnarray}{Cl} 
           \mathbf{D} &=    \sumK w_k\mathbf{C}_k. 
           \label{eq:computation_of_D}
\end{IEEEeqnarray}
Then, $g_2$ is a minorizing function of $f$ at the points $\PoneIterd,\PtwoIterd, \cdots, \PKIterd$.
\end{theorem}
\begin{IEEEproof}
The proof is provided in Appendix \ref{sec:proof_of_minorizing_function_2}. 
\end{IEEEproof}

The minorizing function $g_2$ provided in 
Theorem \ref{th:minorizing_function_2} is simliar to $g_1$. It is also a concave function of the precoders.  Now, $\mathbf{D}$ is the same for all the precoders, which is very useful in reducing the complexity. 

\subsection{Proposed Algorithms}
With the proposed two minorizing functions, we are able to derive algorithms based on the MM algorithm. 
In the following, we first introduce the MM algorithm which can be used to obtain a stationary point of the optimization problem.

Let $f$ denote the objective function $\sum_{k=1}^K w_k\calRk^{ub}$ in the optimization problem
\eqref{eq:objective_optimization_problem_upper_bound}, and $g$ is a minorizing function of it.  
The MM algorithm is an iterative algorithm.
When both $g$ and $f$ are continuously differentiable with respect to the precoding vectors, we have
    \begin{IEEEeqnarray}{Cl}
        &\left.\frac{\partial g}{\partial \Pk^*}\right|_{\Pk=\PkIterd}
        =\left.\frac{\partial f}{\partial \Pk^*}\right|_{\Pk=\PkIterd}, k=1, \cdots, K.
        \label{eq:minorize_condition_3}
    \end{IEEEeqnarray}

In comparison with the original objective function, the surrogate function $g$ should be easier to optimize.
When a minorizing function is found, it will be maximized instead of the original function. The precoders $\Pone^{(d+1)},\Ptwo^{(d+1)}, \cdots, \PK^{(d+1)}$ at the next iteration is obtained as the maximal point of
$g$ under the constraint.
From equation \eqref{eq:minorize_condition_1}, we obtain that
 \begin{IEEEeqnarray}{Cl}
        &f(\Pone^{(d+1)},\Ptwo^{(d+1)}, \cdots, \PK^{(d+1)})  \nonumber \\
        &~~\geq f(\PoneIterd,\PtwoIterd, \cdots, \PKIterd).
        \label{eq:monotone_of_iterations}
    \end{IEEEeqnarray}
Based on equations \eqref{eq:minorize_condition_3} and \eqref{eq:monotone_of_iterations}, the sequence will converge to a stationary point
of the original function $f$. The rigorous proof of the convergence of the MM algorithms depends on the condition \eqref{eq:minorize_condition_1}  and can be found in the literature \cite{vaida2005parameter,jacobson2007expanded}. 

Based on the minorizing function $g_1$ and the MM algorithm, we  obtain 
a sequence of precoding vectors  by
\begin{IEEEeqnarray}{Cl}
        \Pone^{(d+1)},\Ptwo^{(d+1)}, \cdots, \PK^{(d+1)}
         &=\argmax{\Pone,\cdots,\PK}g_1(\Pone,\Ptwo, \cdots, \PK)  \nonumber \\
        &~~{\mathrm s.t.} ~  \sumK \tr{\Pk\Pk^H} \leq 1.
        \label{eq:update_through_surrogate_optimization_problem_1}
\end{IEEEeqnarray}
The sequence of the precoders provided by \eqref{eq:update_through_surrogate_optimization_problem_1} converges to a stationary point of \eqref{eq:objective_optimization_problem_upper_bound}.
In the optimization problem  \eqref{eq:update_through_surrogate_optimization_problem_1}, the precoders of different users are decoupled. It makes the surrogate problem easier to optimize than the original problem. Furthermore, the surrogate problem is a classic concave quadratic optimization problem. For such problems, the optimal solution can be found by using the Lagrange
multiplier method. Let the Lagrangian of the optimization problem \eqref{eq:update_through_surrogate_optimization_problem_1} be defined as
\begin{IEEEeqnarray}{Cl}
        &\mathcal{L}(\mu, \Pone,\Ptwo, \cdots, \PK) 
        \nonumber \\
        &~~~~= - g_1+ \mu(\sumK \tr{\Pk\Pk^H} - 1)
        \label{eq:lagrangian_of_surrogate_problem_1}
\end{IEEEeqnarray}
where $\mu$ is the Lagrange multiplier.
From the first order optimal conditions of \eqref{eq:lagrangian_of_surrogate_problem_1}, we obtain that
\begin{IEEEeqnarray}{Cl}
              \Pk^{(d+1)} &= (\DkIterd+\mu^\star\IMt)^{-1}w_k\AkIterd\PkIterd.
              \label{eq:optimal_solution_of_surrogate_optimization_1}
\end{IEEEeqnarray}
The total power $\sum_{k=1}^K{\mathrm{tr}}(\Pk\Pk^H)$ is a monotonically decreasing function of $\mu$. Thus, the optimal 
$\mu^\star$ is easy to obtain.
If $\mu^\star=0$ and $\sum_{k=1}^K{\mathrm{tr}}(\Pk^{(d+1)}(\Pk^{(d+1)})^H)  \leq 1$, the optimal solution $\Pk^{(d+1)}=(\DkIterd)^{-1}w_k\AkIterd\PkIterd$ is already obtained. Otherwise, we can obtain the optimal multiplier $\mu^\star$ by using a bisection method.

We now summarize the algorithm for the design of the linear precoder by using $g_1$.

\vspace{0.5em}
\hrule
\vspace{0.5em}
Algorithm 1
\vspace{0.5em}
\hrule
\begin{enumerate}[\IEEElabelindent=3em]
\item[Step 1:]
Set $d=0$. Initialize $\Pone^{(d)}$, $\Ptwo^{(d)}$, $\cdots$, $\PK^{(d)}$ and normalize them to satisfy the power constraint.
\item[Step 2:]
Calculate $v_k^{(d)}$ and $\check{v}_k^{(d)}$ according to
\begin{IEEEeqnarray}{Cl}
        v_k^{(d)}  
        &=\sz +  \sumnok (\Pl^{(d)})^H\mathbf{R}_k \Pl^{(d)}  \nonumber
        \\
	\check{v}_k^{(d)} &= v_k^{(d)}+   (\Pk^{(d)})^H\mathbf{R}_k\Pk^{(d)}. \nonumber  
\end{IEEEeqnarray}
\item[Step 3:]
Compute $\mathbf{A}_k^{(d)}$ and ${\mathbf{D}}_k^{(d)}$ according to
\begin{IEEEeqnarray}{Cl}
        \mathbf{A}_k^{(d)}  &= (v_k^{(d)})^{-1}\mathbf{R}_k. \nonumber 
        \\   
           \mathbf{B}_k^{(d)} &=     (v_k^{(d)})^{-1}(\check{v}_k^{(d)})^{-1}\mathbf{R}_k\Pk^{(d)}(\Pk^{(d)})^H\mathbf{R}_k \nonumber
           \\
            \mathbf{C}_k^{(d)} &=    ( (v_k^{(d)})^{-1}- (\check{v}_k^{(d)})^{-1})\mathbf{R}_k  \nonumber
          \\
           \mathbf{D}_k^{(d)}&= w_k\mathbf{B}_k^{(d)} + \sumnok w_l\mathbf{C}_l^{(d)}.  \nonumber
\end{IEEEeqnarray}
    
\item[Step 4:]
Update $\mathbf{p}_{k}^{(d+1)}$ by
\begin{IEEEeqnarray}{Cl}
              \Pk^{(d+1)} &= (\mathbf{D}_k^{(d)}+\mu^\star\IMt)^{-1}w_k\AkIterd\PkIterd  \nonumber 
\end{IEEEeqnarray}
where 
 $\mu^\star$ is computed by the bisection method. 
Set $d=d+1$.
\end{enumerate}
\hspace{1em}Repeat Step 2 through Step 4 until convergence or until a
pre-set target is reached.
\vspace{0.5em}
\hrule
\vspace{0.5em}

The above algorithm relies on the posterior beam based statistical channel model in  \eqref{eq:posterori_model_of_Hkm} through the computation of the matrix $\mathbf{R}_k$. It can also apply to other channel models by changing the computation of $\mathbf{R}_k$ accordingly.
In Step 4, the matrices $\DkIterd$ are different for different users.  Thus, we need to perform a matrix inversion $(\DkIterd+\mu^\star\IMt)^{-1}$ for each user. For large $M_t$, the complexity of the matrix inversion is very high. The complexity can be reduced if the number of the matrix inversions is reduced. If the matrices $\DkIterd$ are the same for all the users as that in the minorizing function $g_2$, we only need one matrix inversion. Thus, we continue to derive another algorithm based on the minorizing function $g_2$.

The steps of using the minorizing function $g_2$ to obtain a
local optimal solution  is the same to that of using $g_1$. For brevity,  the details are omitted here. The solution is given directly as
\begin{IEEEeqnarray}{Cl}
              \Pk^{(d+1)} &= (\mathbf{D}^{(d)}+\mu^\star\IMt)^{-1}w_k\AkIterd\PkIterd.
              \label{eq:optimal_solution_of_surrogate_optimization_2}
\end{IEEEeqnarray}
Now, only one matrix inversion is needed for the computation of all the precoders.
 
 For comparison, we write the iterative process of updating  the precoders  as
 \begin{IEEEeqnarray}{Cl}
             \Pk^{(d+1)} &= (\sumK w_k ((v_k^{(d)})^{-1}-(\check{v}_k^{(d)})^{-1})\mathbf{R}_k+\mu^\star\IMt)^{-1}\nonumber \\
            &~~~~~~w_k (v_k^{(d)})^{-1}\mathbf{R}_k\Pk^{(d)}.
              \label{eq:optimal_solution_of_surrogate_optimization_iterative}
\end{IEEEeqnarray}
The iteration process looks quite simple. The proposed precoder in \eqref{eq:optimal_solution_of_surrogate_optimization_iterative} is for massive MIMO with single antenna users. A similar precoder can also be obtained for massive MIMO with multiple antenna users, however, the rigorous mathematic proof for its convergence might not be easy to obtain.
 The weight $\gamma_k=v_k^{-1}-\check{v}_k^{-1}$ in  \eqref{eq:optimal_solution_of_surrogate_optimization_iterative}  is very interesting, we discuss it further in the following. It can be rewritten as 
 \begin{IEEEeqnarray}{Cl}
\gamma_k &=  v_k ^{-1} \check{v}_k ^{-1} \Pk ^H\mathbf{R}_k \Pk  \nonumber \\
 &=\check{v}_k^{-1}\eta_k
 \label{eq:gamma_k}
 \end{IEEEeqnarray}
 where $\eta_k={v}_k^{-1}\Pk^H\mathbf{R}_k \Pk$ can be viewed as the signal to interference plus noise ratio (SINR) of the $k$-th user. From  \eqref{eq:gamma_k}, we obtain that the weight $\gamma_k$ is proportional to the SINR and inversely proportional to the total receive energy $\check{v}_k$. 
Thus, it represents the sensitivity of the channel of the $k$-th user to the interference. If we can know these weights before precoding by using the machine learning method, we are able to further reduce the complexity of the proposed precoder. 

When the CSI is perfectly known at the BS, the upper bound of the expected rate used in this subsection will become the exact rate.
 In such case, the iterative formula of the precoder obtained in \eqref{eq:optimal_solution_of_surrogate_optimization_iterative} becomes
 \begin{IEEEeqnarray}{Cl}
       &   \mathbf{p}_k^{(d+1)}  = (\sumK w_k((v_k^{(d)})^{-1}-(\check{v}_k^{(d)})^{-1})\mathbf{h}_{k}\mathbf{h}_{k}^H+ \mu^*\mathbf{I})^{-1}
         \nonumber \\
         &~~~~~~~~ w_k(v_k^{(d)})^{-1}\Hk\Hk^H\mathbf{p}_k^{(d)}
        \label{eq:low_complexity_robust_precoder_design_perfect_CSI}
\end{IEEEeqnarray}
 which is equivalent to the iterative WMMSE precoder.
  The RZF and SLNR precoders are two widely used precoders in the literature.   
 For the considered massive MIMO, the RZF precoder is given by \cite{joham2002transmit}
\begin{IEEEeqnarray}{Cl}
    \mathbf{p}_{k}^{\rm RZF} =  \rho (\sumK  \mathbf{h}_{k}\mathbf{h}_{k}^H+K\sigma_z^2\mathbf{I})^{-1}\mathbf{h}_k
\end{IEEEeqnarray} 
where  $\rho$ is the power normalized factor.
Let 
$P_k$ denote the power constraint of the $k$-th user. The SLNR precoder is given by \cite{sadek2007leakage} 
\begin{IEEEeqnarray}{Cl}
    &\mathbf{p}_{k}^{\rm SLNR}  \propto  \max \nonumber \\
    &~~~~  {\rm eigenvector}\left\{(\sumnok\mathbf{h}_l\mathbf{h}_l^H+\frac{1}{P_k}\sigma_z^2\mathbf{I})^{-1}\mathbf{h}_k\mathbf{h}_k^H \right\}.
    \label{eq:pslnr}
\end{IEEEeqnarray}  
Since the iterative WMMSE precoder can achieve better performance than that of the RZF and SLNR precoders, the proposed precoder is also superior to these two precoders.

For the cases where $\Hk^H$ are not perfectly known at the BS, 
the proposed precoder utilizes both the instantaneous and statistical CSI to reduce interference, and thus achieves better performance than the precoders that only use inaccurate instantaneous CSI. 
To get more insights, we consider an extreme case when $\Hk$ has zero means.
In such case, the instantaneous CSI is useless for the precoder design.
Thus, the WMMSE, RZF, and SLNR precoders that depend on the instantaneous CSI can not work. 
For the proposed precoder, $\mathbf{A}_k$ becomes the weighted channel covariance matrix of the $k$-th user, and  $\mathbf{D}  + \mu \mathbf{I}$  is dominated by the weighted channel covariance matrices of the interference users. 
Base on the posterior channel model, we can still obtain the precoders that simultaneously guarantee the gains of the signal and keep the interference small.

We now summarize the algorithm for the design of precoder by utilizing $g_2$. 
\vspace{0.5em}
\hrule
\vspace{0.5em}
Algorithm 2 
\vspace{0.5em}
\hrule
\begin{enumerate}[\IEEElabelindent=3em]
\item[Step 1:]
Set $d=0$. Initialize $\Pone^{(d)}$, $\Ptwo^{(d)}$, $\cdots$, $\PK^{(d)}$ and normalize them to satisfy the power constraint.
\item[Step 2:]
Calculate $v_k^{(d)}$ and $\check{v}_k^{(d)}$ according to
\begin{IEEEeqnarray}{Cl}
        v_k^{(d)}  
        &=\sz +  \sumnok (\Pl^{(d)})^H\mathbf{R}_k \Pl^{(d)}  \nonumber
        \\
	\check{v}_k^{(d)} &= v_k^{(d)}+   (\Pk^{(d)})^H\mathbf{R}_k\Pk^{(d)}. \nonumber  
\end{IEEEeqnarray}
\item[Step 3:]
Compute $\mathbf{A}_k^{(d)}$ and ${\mathbf{D}}^{(d)}$ according to
\begin{IEEEeqnarray}{Cl}
        \mathbf{A}_k^{(d)}  &= (v_k^{(d)})^{-1}\mathbf{R}_k. \nonumber 
        \\  
          \mathbf{C}_k^{(d)} &=    ( (v_k^{(d)})^{-1}- (\check{v}_k^{(d)})^{-1})\mathbf{R}_k.  \nonumber
           \\
           \mathbf{D}_k^{(d)}&= \sumK w_k\mathbf{C}_k^{(d)}.  \nonumber
\end{IEEEeqnarray}
    
\item[Step 4:]
Update $\mathbf{p}_{k}^{(d+1)}$ by
\begin{IEEEeqnarray}{Cl}
              \Pk^{(d+1)} &= (\mathbf{D}^{(d)}+\mu^\star\IMt)^{-1}w_k\AkIterd\PkIterd  \nonumber 
\end{IEEEeqnarray}
where 
 $\mu^\star$ is computed by the bisection method. 
Set $d=d+1$.
\end{enumerate}
\hspace{1em}Repeat Step 2 through Step 4 until convergence or until a
pre-set target is reached.
\vspace{0.5em}
\hrule
\vspace{0.5em}
Similarly, Algorithm 2 also relies on the posterior beam based statistical channel model through the computation of the matrix $\mathbf{R}_k$. It can also be used in other channel models by changing the computation of $\mathbf{R}_k$ accordingly.

\subsection{Complexity Analysis}

For very large $M_t$, the complexity of Algorithm 1 is dominated by the matrix inversion and the product of the matrix and the vector in Step 4.
Since the complexity of the matrix inversion is $M_t^3/2$ and the complexity of the product of the matrix and the vector is $KM_t^2$, we have
the complexity of Step 4 is $KM_t^3/2+2KM_t^2$. 
Thus, the complexity of Algorithm 1 is of order $\mathcal{O}(KM_t^3/2+2KM_t^2)$ per iteration.
The complexity of Algorithm 2 is also dominated by Step 4. However, only one matrix inversion is needed for Algorithm2. Thus, its complexity is of order $\mathcal{O}(M_t^3/2+2KM_t^2)$ per iteration, which is obviously lower than that of Algorithm 1.
The SLNR precoder and the RZF precoder has nearly the same complexity since the equivalence between them under equal power allocation \cite{patcharamaneepakorn2012equivalence}. 
Thus, we only present the complexity of the RZF precoder for comparison.
For the RZF precoder, the complexity is $\mathcal{O}(K^3/2+2K^2M_t)$, 
and we have that the complexity of Algorithm 2 per iteration is similar to that of the RZF precoder when $K$ is also very large.
For the case where $K$ is not very large,
the truncated conjugate gradient (CG) method or the matrix inversion lemma can be used to further reduce the complexity of Algorithm 2.

\begin{figure}[h]
\centering
\includegraphics[scale=0.45]{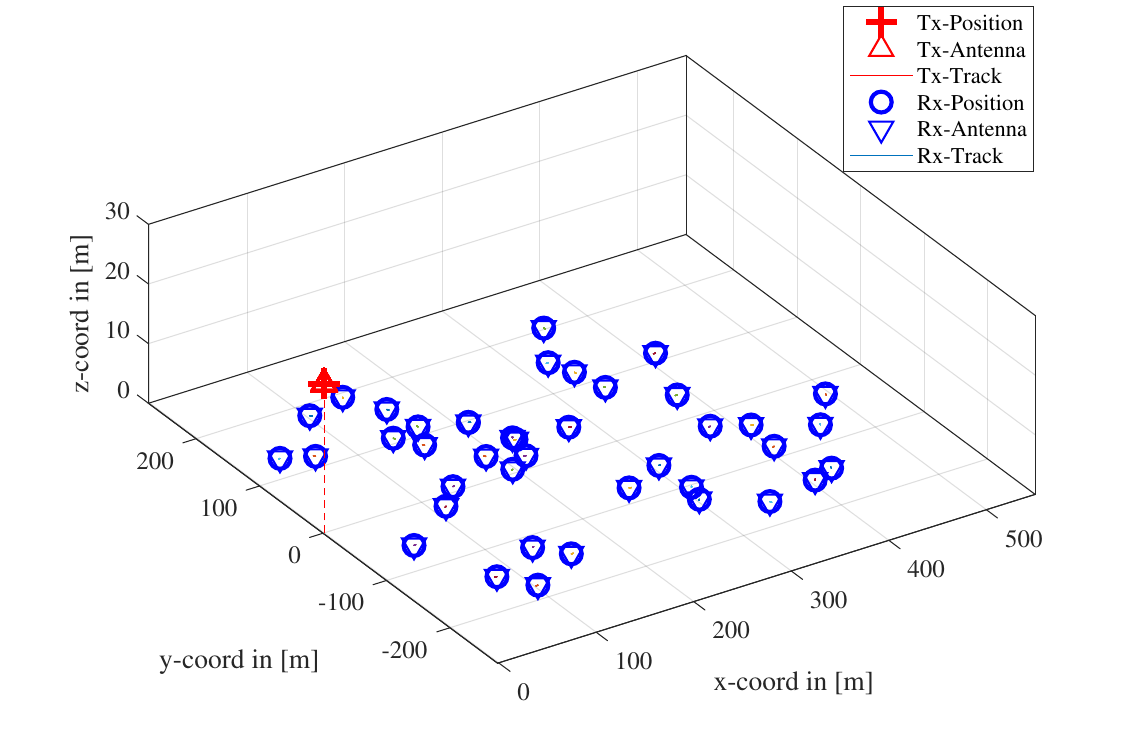}
\caption{The layout of a massive MIMO with $M_t=128, M_r=1$, $K=40$.}
\label{fig:Layout}
\end{figure}
\section{Simulation Results}
In this section, we provide simulation results to show the performance of the proposed precoder. We use the widely used QuaDRiGa channel model \cite{jaeckel2014quadriga}.
For simplicity, the path loss model and shadow fading are disabled.
We set the center frequency to $4.8$ GHz.
The simulation scenario is set to ``3GPP\_38.901\_UMa\_NLOS''.
The total transmit power is set as $P=1$.
The type of the antenna array used at the BS is ``$\rm{3gpp}$-$\rm{3d}$'' with $D_{in}=1$.
We consider a massive MIMO with $M_t=128$ antennas at the BS, where $M_x=16$ and $M_z=8$. The number of users is set as
$K=40$, and each user is equipped with single antenna. We only use the setting $N_t \geq M_t$ in the BS side.
The layout of this massive MIMO system is plotted in Fig.~\ref{fig:Layout}, where
the location of the BS is at $(0,0,25)$,
and the users are randomly generated in a circle with radius $r=250$m around $(0,0,0)$ at $1.5$m height and then move $250$m to the right side of the BS.
For simplicity, the SNR is set as SNR=$\frac{1}{\sz}$.
The lengths of all the simulated tracks are set to be the same $2$ m.

\begin{figure}[h]
\centering
\includegraphics[scale=0.6]{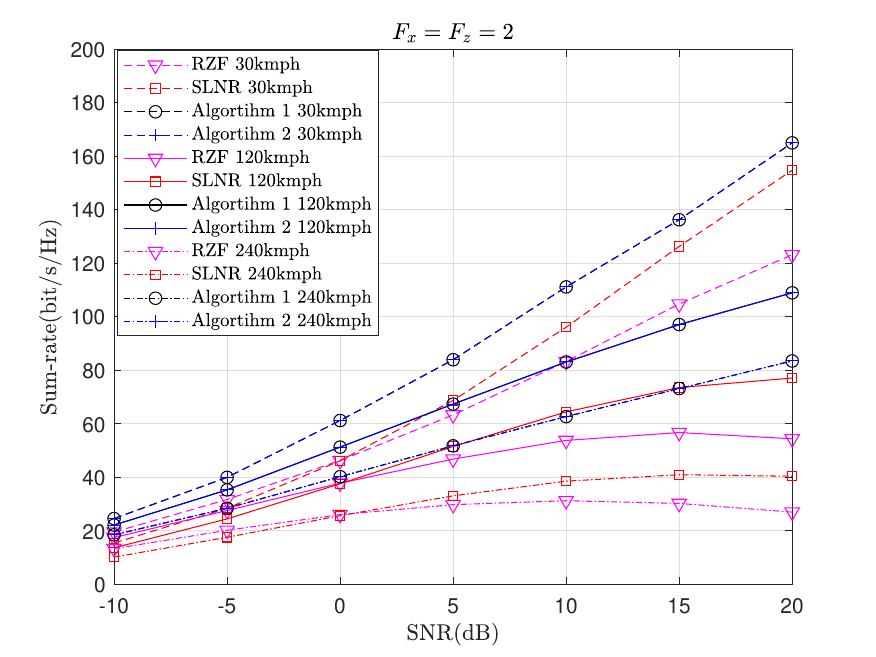}
\caption{The sum-rate performance of  four precoders for a massive MIMO downlink with $M_t=128, M_r=1$, $K=40$ and $F_x =F_z = 2$.}
\label{fig:Capacity_64Tx_9user_1Rx_NLOS_5_Det}
\end{figure}

To use the proposed robust precoders, we first compute  $\alpha_{kn}$ as the empirical  temporal correlation coefficients through 
\begin{equation}
\alpha_{kn}=\frac{1}{M_sN_b}\sum_{m,i}\frac{{\rm abs}({\rm tr}(\mathbf{h}_{kmi}\mathbf{h}_{km(i+n)}^*))}{\sqrt{{\rm tr}(\mathbf{h}_{kmi}\mathbf{h}_{kmi}^*){\rm tr}(\mathbf{h}_{km(i+n)}\mathbf{h}_{km(i+n)}^*)}}
\end{equation}
where $\mathbf{h}_{km(i+n)}$ denotes the channel whose location is $n$  blocks after the $i$-th block on the $m$-th subframe, and $M_s$ is the number of used subframes. We then obtain the channel power matrices $\boldsymbol{\omega}_k$  from the sample of channel matrices.
With the channel power matrices $\boldsymbol{\omega}_k$ and the temporal correlation coefficient $\alpha_{kn}$, we are able to perform Algorithms 1 and 2.
Fig.~\ref{fig:Capacity_64Tx_9user_1Rx_NLOS_5_Det} plots the sum-rate performance of Algorithms 1 and 2, the RZF precoder   and the SLNR precoder for the considered massive MIMO downlink.
The length of one subframe is set to $1$ ms. The number of the blocks is set as $N_b=7$.
The imperfect CSI was caused by channel aging, which was related to the moving speeds of the users.
To show the impact of imperfect CSI, the speed of the users are set to $30$, $120$ and $240$ km/h.
The fine factors of sampled directional cosines are set as $F_x=F_z=2$. The number of iterations is $20$.
We use the RZF precoders as the initial values of Algorithms 1 and 2.
From Fig.~\ref{fig:Capacity_64Tx_9user_1Rx_NLOS_5_Det}, we observe that the performance of Algorithms 1 and 2 are almost the same, and
Algorithm 2 outperforms the RZF precoder and the SLNR precoder significantly at all three cases.
The sum-rate of Algorithm 2 is about $1.34$ times of that of the RZF precoder at SNR=$20$dB for the $30$kmph case. It increases to $2.01$ and $3.07$ times of that of the RZF precoder for the latter two cases.
Meanwhile, the sum-rate of Algorithm 2 is about $1.07$, $1.41$ and $2.09$ times of that of the SLNR precoder for the $30$, $120$ and $240$ km/h cases.
The results show that the performance gain of robust linear precoders are more significant in high mobility scenario.

\begin{figure}[h]
\centering
\includegraphics[scale=0.6]{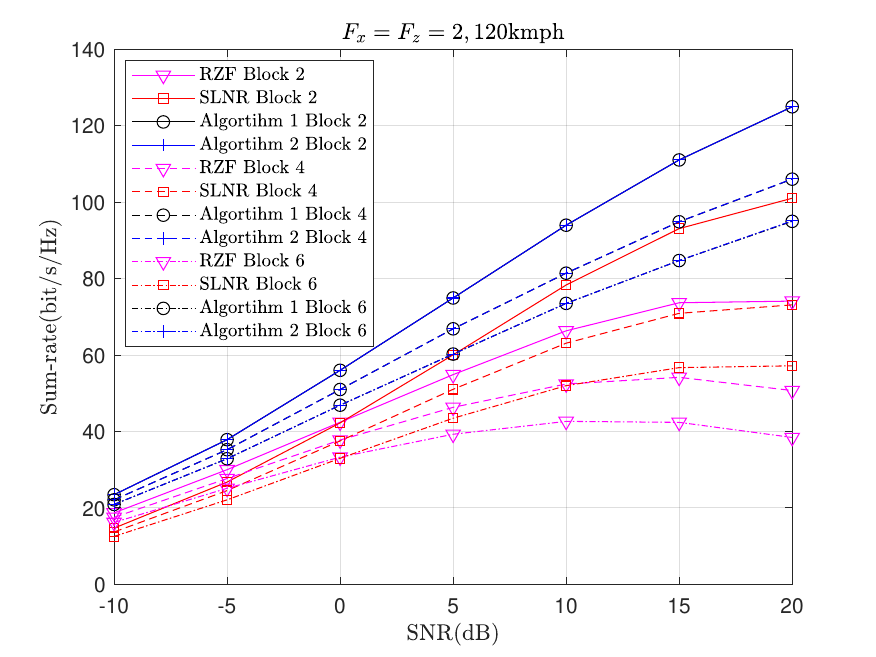}
\caption{The sum-rate performance of four precoders over different blocks for a massive MIMO downlink with $M_t=128, M_r=1$, $K=40$, $N_b=7$ and $F_x =F_z = 2$.}
\label{fig:Capacity_128Tx_40users_1Rx_Blocks}
\end{figure}
Simulation results provided in Fig.~\ref{fig:Capacity_64Tx_9user_1Rx_NLOS_5_Det} are the sum-rate performance of the considered massive MIMO for each subframe, where the sum-rates of different blocks have been averaged. To get more details about the performance of the proposed precoders, we fix the speed of the users to  $120$ km/h, and provides the sum-rate performance of four precoders over different blocks for the considered massive MIMO downlink in Fig.~\ref{fig:Capacity_128Tx_40users_1Rx_Blocks}. The temporal correlation coefficient $\alpha_{kn}$ becomes smaller as the block number increases. As shown in Fig.~\ref{fig:Capacity_128Tx_40users_1Rx_Blocks}, 
 the performance gains of the proposed precoders on each block become more significant when $\alpha_{kn}$ is smaller or SNR is higher. It  shows the degradations of the proposed precoders are much smaller compared to the other two precoders as $\alpha_{kn}$ decreases, and they can solve the problem of channel aging. When the duration of one subframe becomes higher, the channel aging effect will also become more severe. The effect of increasing the duration of one subframe is similar to that of  increasing the speeds of the users simultaneously. Thus, the sum-rate performance of the proposed precoders on each subframe with higher duration will be similar to that of the case with higher speed provided in  Fig.~\ref{fig:Capacity_64Tx_9user_1Rx_NLOS_5_Det}. In other words, the sum-rate performance  of all precoders will degrade in the case with higher duration, but the degradations of the proposed precoders are much smaller than that of the SLNR and RZF precoders, which means the performance gain of the proposed precoders will also be more significant.

\begin{figure}[h]
\centering
\includegraphics[scale=0.55]{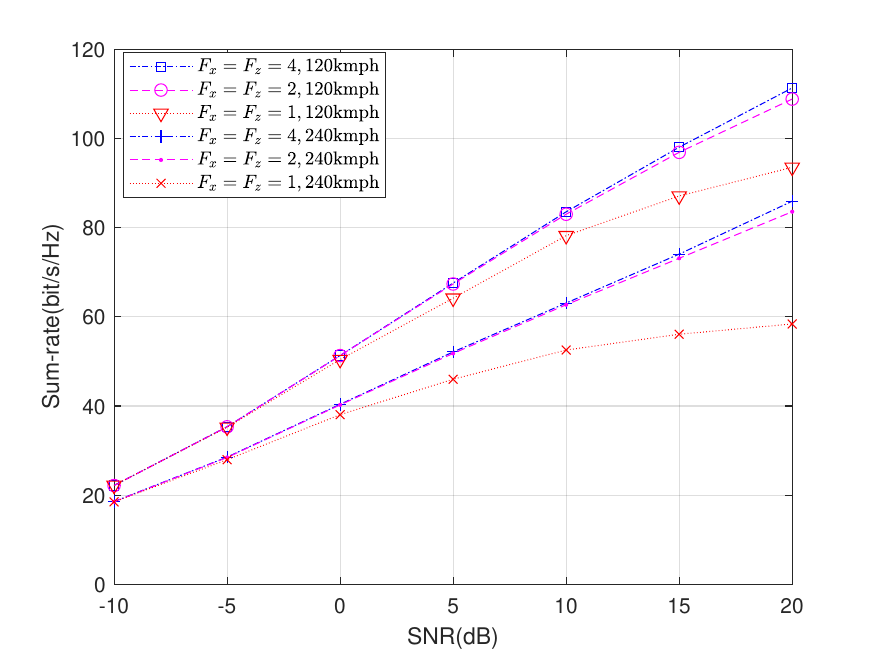}
\caption{The sum-rate performance of the low complexity robust linear precoders with different fine factors for a massive MIMO downlink with $M_t=128, M_r=1$, $K=40$.}
\label{fig:Capacity_128Tx_16user_1Rx_NLOS_5_three_os_factors}
\end{figure}

In the previous simulations, we have set the fine factors of steering vectors at the BS side as $F_x=F_z=2$.
To investigate the impacts of  the fine factors $F_x$ and $F_z$, we simulate the sum-rate of Algorithm 2 for three cases: case one where $F_x=F_z=4$, case two where $F_x=F_z=2$  and case three where $F_x=F_z=1$.
The length of one subframe is still set to $1$ ms.
We consider both the moderate and high mobility scenario. The users' speed is set to $120$ and $240$km/h.
Simulation results of the sum-rates are shown in Fig.~\ref{fig:Capacity_128Tx_16user_1Rx_NLOS_5_three_os_factors}.
It can be observed that using
large $F_x$ and $F_z$ achieves better performance.
The sum-rate of the $F_x=F_z=2$ case is about $1.16$ times of that with $F_x=F_z=1$ at SNR=$20$dB for the moderate speed scenario. It increases to $1.43$ times for the high mobility scenario.
Since the established channel model with $F_x=F_z=1$ is equivalent to the beam domain channel model, 
it shows that the beam domain channel model is not accurate enough for the considered massive MIMO with UPA, and the robust precoder designed by using the beam based statistical  channel model is superior.
Furthermore, the sum rate of $F_x = F_z = 1$ cases increases more  slowly when SNR is higher. This is because the part of error caused by the channel model gradually becomes dominant  as the SNR increases.
Finally, the performance gain of  the robust precoder with   $F_x=F_z=4$ is not significant compared to that of $F_x=F_z=2$. Thus, to achieve a good precoding performance, the established beam based statistical channel model with $F_x=F_z=2$ is accurate enough.

\begin{figure}[h]
\centering
\includegraphics[scale=0.55]{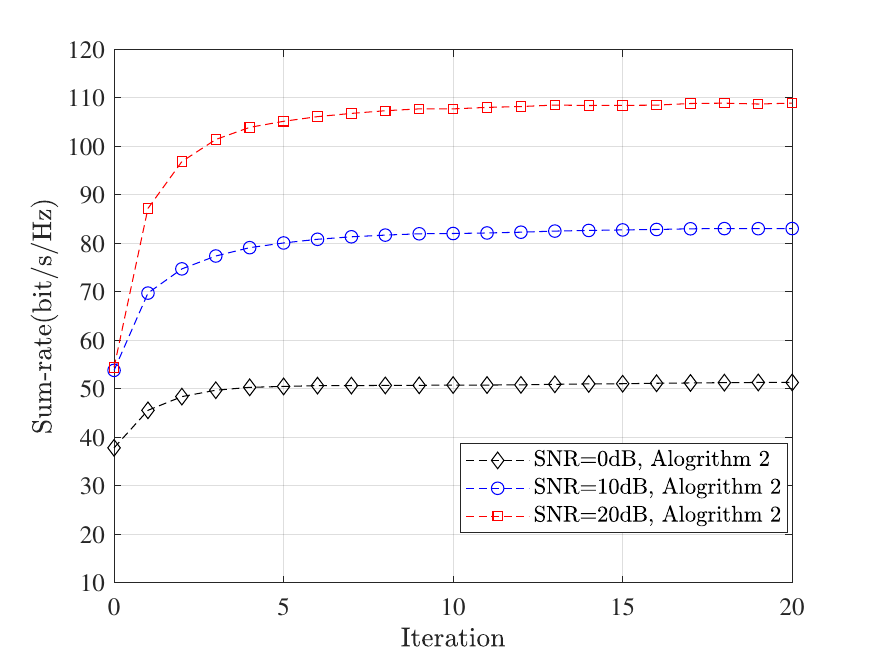}
\caption{The convergence behavior of Algorithm 2 at each iteration for three  different SNRs.}
\label{fig:Convergence_behaviour_of_128Tx_16user_1Rx_NLOS_5}
\end{figure}
We then study the convergence behavior of Algorithm 2.
The speed of the users is set to $120$ km/h.
We still use the RZF precoders as the initial values.
The number of the iterations is $20$.
Fig.~\ref{fig:Convergence_behaviour_of_128Tx_16user_1Rx_NLOS_5}
plots the sum-rates of the proposed precoder at each iteration for three different SNRs.
From Fig.~\ref{fig:Convergence_behaviour_of_128Tx_16user_1Rx_NLOS_5}, we see that
the proposed algorithm for all three cases quickly converges.
We also observe that it takes more iterations to converge as the SNR increases. At SNR= $0$dB, only $5$
iterations are needed for the convergence, whereas $10$ iterations are needed at SNR= $20$dB.
The number of iterations can be further reduced if we use better initial values such as the precoder from previous subframe.

\section{Conclusion}
In this paper, we investigated the robust linear precoder design for massive MIMO downlink with UPA and imperfect CSI, which was represented by an \textit{a posteriori} beam based statistical channel model. The posterior channel model is established from a  \textit{a priori} beam based statistical channel model based on the matrix of sampled steering vectors.
On the basis of the posterior channel model, the considered optimization problem of robust precoder design was maximizing an upper bound of the expected weighted sum-rate under a total power constraint. 
By deriving two minorizing functions and using the MM algorithm, we derived two iterative algorithms to obtain the local optimal precoders.
Simulation results showed that the proposed precoders can achieve significantly performance gain compared to the RZF precoder and the SLNR precoder.

\appendices
\section{Proof of Theorem \ref{th:minorizing_function_1}}
\label{sec:proof_of_minorizing_function_1} 
Recall that the upper bound of the expected rate can be written as
  \begin{IEEEeqnarray}{Cl}
          \calRk^{ub}  =  \log({ 1 + v_k^{-1}\mean{\Hk^H\Pk\Pk^H\Hk}}). 
    \end{IEEEeqnarray}
Since $\Pk^H\Hk$ is a scalar, we have $\Hk^H\Pk\Pk^H\Hk=\Pk^H\Hk\Hk^H\Pk$. Then, we obtain that
\begin{IEEEeqnarray}{Cl}
          \calRk^{ub}  =  \log({ 1 + v_k^{-1}\Pk^H\mean{\Hk\Hk^H}}\Pk). 
\end{IEEEeqnarray}
Let $e_k$ be defined as
\begin{IEEEeqnarray}{Cl}
          e_k = ( 1 + v_k^{-1}\Pk^H\mean{\Hk\Hk^H}\Pk)^{-1}. 
\end{IEEEeqnarray} 
We then have 
\begin{IEEEeqnarray}{Cl}
               \calRk^{ub}  =  -\log(e_k).
\end{IEEEeqnarray} 
Since $-\log(\cdot)$ is a convex function, we have that
\begin{IEEEeqnarray}{Cl}
               \calRk^{ub}  \ge  -\log(e_k^{(d)})  - (e_k^{(d)})^{-1}(e_k-e_k^{(d)})
\end{IEEEeqnarray}
where the equality holds when $e_k=e_k^{(d)}$.
By defining $a_k=-\log(e_k^{(d)}) + 1$, we obtain that
\begin{IEEEeqnarray}{Cl}
               \calRk^{ub}  \ge  a_k  - (e_k^{(d)})^{-1}e_k. 
               \label{eq:calRk_lower_bound_one}
\end{IEEEeqnarray}
From $\mathbf{R}_k = \mean{\Hk\Hk^H}$, we rewrite $e_k$   as
\begin{IEEEeqnarray}{Cl}
          e_k = ( 1 + v_k^{-1}\Pk^H\mathbf{R}_k \Pk)^{-1}. 
\end{IEEEeqnarray} 
The function $e_k$ is still complicated since the precoder $\Pk$ and those in $v_k$ are coupled. Thus, we need to find a way to decouple them. Inspired by \cite{shi2011iteratively}, we notice that $e_k $ is also the minimum of the optimization problem 
\begin{IEEEeqnarray}{Cl}
          \min_{\mathbf{g}_k}  ( 1 - \mathbf{g}_k^H\mathbf{R}_k^{1/2}\mathbf{p}_k)( 1 - \mathbf{g}_k^H\mathbf{R}_k^{1/2}\mathbf{p}_k )^*+v_k\mathbf{g}_k^H\mathbf{g}_k. 
\end{IEEEeqnarray} 
Thus, for any $\mathbf{g}_k$, we obtain that  
\begin{IEEEeqnarray}{Cl}
            (e_k^{(d)})^{-1}e_k  &\le     (e_k^{(d)})^{-1}( 1 - \mathbf{g}_k^H\mathbf{R}_k^{1/2}\mathbf{p}_k)( 1 - \mathbf{g}_k^H\mathbf{R}_k^{1/2}\mathbf{p}_k)^* \nonumber \\
            &~~~~ + (e_k^{(d)})^{-1}v_k\mathbf{g}_k^H\mathbf{g}_k
\end{IEEEeqnarray} 
where the equality holds when 
\begin{IEEEeqnarray}{Cl}
          \mathbf{g}_k^H =v_k^{-1} \Pk^H\mathbf{R}_k^{1/2}(1 +  v_k^{-1}\mathbf{R}_k^{1/2}\Pk\Pk^H\mathbf{R}_k^{1/2})^{-1} \nonumber \\
     =v_k^{-1}(1 +  v_k^{-1} \Pk^H\mathbf{R}_k\Pk)^{-1}\Pk^H\mathbf{R}_k^{1/2}. 
\end{IEEEeqnarray} 
We define
\begin{IEEEeqnarray}{Cl}
         & (\mathbf{g}_k^{(d)})^H \nonumber \\
          &=(v_k^{(d)})^{-1} (\Pk^{(d)})^H\mathbf{R}_k^{1/2} \nonumber \\
          &~~~~(\mathbf{I} +  (v_k^{(d)})^{-1}\mathbf{R}_k^{1/2}\Pk^{(d)}(\Pk^{(d)})^H\mathbf{R}_k^{1/2})^{-1} \nonumber \\
    & =(v_k^{(d)})^{-1}(1 +  (v_k^{(d)})^{-1} (\Pk^{(d)})^H\mathbf{R}_k\Pk^{(d)})^{-1}(\Pk^{(d)})^H\mathbf{R}_k^{1/2}.  \nonumber \\
\end{IEEEeqnarray} 
Then, we have that
\begin{IEEEeqnarray}{Cl}
           & (e_k^{(d)})^{-1}e_k \nonumber \\
              &\le     (e_k^{(d)})^{-1}( 1 - (\mathbf{g}_k^{(d)})^H\mathbf{R}_k^{1/2}\mathbf{p}_k)( 1 - (\mathbf{g}_k^{(d)})^H\mathbf{R}_k^{1/2}\mathbf{p}_k)^*\nonumber 
            \\
            &~~~~ + (e_k^{(d)})^{-1}v_k(\mathbf{g}_k^{(d)})^H\mathbf{g}_k^{(d)}
\end{IEEEeqnarray} 
where the equality holds when $\mathbf{p}_k=\mathbf{p}_k^{(d)}$ and $v_k = v_k^{(d)}$. From the above equation and \eqref{eq:calRk_lower_bound_one}, we obtain that
\begin{IEEEeqnarray}{Cl}
           \calRk^{ub}  &\ge  a_k  -   (e_k^{(d)})^{-1}+  (e_k^{(d)})^{-1}(\mathbf{g}_k^{(d)})^H\mathbf{R}_k^{1/2}\mathbf{p}_k \nonumber \\
           &~~~~ +  (e_k^{(d)})^{-1}\mathbf{p}_k^H\mathbf{R}_k^{1/2}\mathbf{g}_k^{(d)} \nonumber \\
           &~~~~-(e_k^{(d)})^{-1}(\mathbf{g}_k^{(d)})^H\mathbf{R}_k^{1/2}\mathbf{p}_k\mathbf{p}_k^H\mathbf{R}_k^{1/2}\mathbf{g}_k^{(d)}\nonumber \\
           &~~~~- (e_k^{(d)})^{-1}v_k(\mathbf{g}_k^{(d)})^H\mathbf{g}_k^{(d)}  
\end{IEEEeqnarray} 
 where the equality also holds when $\mathbf{p}_k=\mathbf{p}_k^{(d)}$ and $v_k = v_k^{(d)}$.
 Substituting the formula of $\mathbf{g}_k^{(d)}$ into $ (e_k^{(d)})^{-1}(\mathbf{g}_k^{(d)})^H\mathbf{R}_k^{1/2}\mathbf{p}_k$, we have that
\begin{IEEEeqnarray}{Cl}
             (e_k^{(d)})^{-1}(\mathbf{g}_k^{(d)})^H\mathbf{R}_k^{1/2}\mathbf{p}_k   = (v_k^{(d)})^{-1}(\mathbf{p}_k^{(d)})^H\mathbf{R}_k \mathbf{p}_k.
             \label{eq:substitute_results_one}
\end{IEEEeqnarray} 
Similarly, we obtain that 
\begin{figure*}
\begin{IEEEeqnarray}{Cl}
          &(e_k^{(d)})^{-1}(\mathbf{g}_k^{(d)})^H\mathbf{R}_k^{1/2}\mathbf{p}_k\mathbf{p}_k^H\mathbf{R}_k^{1/2}\mathbf{g}_k^{(d)} \nonumber \\
          &=  (v_k^{(d)})^{-1}(\mathbf{p}_k^{(d)})^H\mathbf{R}_k \mathbf{p}_k\mathbf{p}_k^H\mathbf{R}_k^{1/2}(v_k^{(d)})^{-1} (\mathbf{I} +  (v_k^{(d)})^{-1}\mathbf{R}_k^{1/2}\Pk^{(d)}(\Pk^{(d)})^H\mathbf{R}_k^{1/2})^{-1} \mathbf{R}_k^{1/2}(\Pk^{(d)}) 
          \nonumber \\
          &=  (v_k^{(d)})^{-1}(\mathbf{p}_k^{(d)})^H\mathbf{R}_k \mathbf{p}_k\mathbf{p}_k^H\mathbf{R}_k^{1/2}  (v_k^{(d)}\mathbf{I}+   \mathbf{R}_k^{1/2}\Pk^{(d)}(\Pk^{(d)})^H\mathbf{R}_k^{1/2})^{-1} \mathbf{R}_k^{1/2}(\Pk^{(d)}) 
          \nonumber \\
          &=   (v_k^{(d)})^{-1}\mathbf{p}_k^H\mathbf{R}_k^{1/2}  (v_k^{(d)}\mathbf{I}+   \mathbf{R}_k^{1/2}\Pk^{(d)}(\Pk^{(d)})^H\mathbf{R}_k^{1/2})^{-1} \mathbf{R}_k^{1/2}(\Pk^{(d)})   (\mathbf{p}_k^{(d)})^H\mathbf{R}_k \mathbf{p}_k
          \nonumber \\
          &=  (v_k^{(d)})^{-1}\mathbf{p}_k^H\mathbf{R}_k^{1/2} \mathbf{R}_k^{1/2}\Pk^{(d) }(v_k^{(d)}+   (\Pk^{(d)})^H\mathbf{R}_k^{1/2}\mathbf{R}_k^{1/2}(\Pk^{(d)}))^{-1}  (\mathbf{p}_k^{(d)})^H\mathbf{R}_k  \mathbf{p}_k
          \nonumber \\
         & =  (v_k^{(d)})^{-1}(\check{v}_k^{(d)} )^{-1}\mathbf{p}_k^H\mathbf{R}_k\Pk^{(d)} (\Pk^{(d)})^H \mathbf{R}_k  \mathbf{p}_k   \label{eq:substitute_results_two}
\end{IEEEeqnarray} 
\end{figure*}
\eqref{eq:substitute_results_two} on the top of the following page, where $\check{v}_k^{(d)} = v_k^{(d)} +   (\Pk^{(d)})^H\mathbf{R}_k\Pk^{(d)}$.
By computing
\begin{IEEEeqnarray}{Cl}
          \mathbf{g}_k^H\mathbf{g}_k  
     &=v_k^{-1}(1 +  v_k^{-1} \Pk^H\mathbf{R}_k\Pk)^{-1}\Pk^H\mathbf{R}_k\Pk^H\nonumber \\
     &~~~~v_k^{-1}(1 +  v_k^{-1} \Pk^H\mathbf{R}_k\Pk)^{-1} 
\end{IEEEeqnarray} 
we then obtain that
\begin{IEEEeqnarray}{Cl}
           &(e_k^{(d)})^{-1}v_k(\mathbf{g}_k^{(d)})^H\mathbf{g}_k^{(d)}  \nonumber \\
           &~~=v_k (v_k^{(d)})^{-1}  (\Pk^{(d)})^H \mathbf{R}_k \Pk^{(d)}    (v_k^{(d)} +   (\Pk^{(d)})^H\mathbf{R}_k\Pk^{(d)})^{-1} \nonumber \\
           &~~= v_k ((v_k^{(d)})^{-1} -  (v_k^{(d)} +   (\Pk^{(d)})^H\mathbf{R}_k\Pk^{(d)})^{-1})
           \nonumber \\
           &~~=v_k ((v_k^{(d)})^{-1} -  (\check{v}_k^{(d)} )^{-1}).\label{eq:substitute_results_three}
\end{IEEEeqnarray} 
According  to equations \eqref{eq:substitute_results_one}, \eqref{eq:substitute_results_two}, \eqref{eq:substitute_results_three}, and 
\begin{equation}
        v_k = \sz +  \sumnok\Pl^H\mathbf{R}_k\Pl   
\end{equation}
we obtain that
\begin{IEEEeqnarray}{Cl}
           \calRk^{ub}  &\ge  a_k  -   (e_k^{(d)})^{-1} +  (v_k^{(d)})^{-1}(\mathbf{p}_k^{(d)})^H\mathbf{R}_k \mathbf{p}_k  \nonumber \\
           &~~+ (v_k^{(d)})^{-1}\mathbf{p}_k^H\mathbf{R}_k\mathbf{p}_k^{(d)} \nonumber \\
           &~~-  (v_k^{(d)})^{-1}(\check{v}_k^{(d)} )^{-1}\mathbf{p}_k^H\mathbf{R}_k\Pk^{(d)} (\Pk^{(d)})^H \mathbf{R}_k  \mathbf{p}_k   \nonumber \\
           &~~-((v_k^{(d)})^{-1} -  (\check{v}_k^{(d)} )^{-1}) (\sz +  \sumnok\Pl^H\mathbf{R}_k\Pl).  
\end{IEEEeqnarray} 
Let the matrices $\mathbf{A}_k$, $\mathbf{B}_k$ and $\mathbf{C}_k$  be defined as
\begin{IEEEeqnarray}{Cl}
        \mathbf{A}_k  &= v_k^{-1}\mathbf{R}_k 
        \\
          \mathbf{B}_k &=    v_k^{-1}\check{v}_k^{-1}\mathbf{R}_k\Pk\Pk^H\mathbf{R}_k 
             \\
          \mathbf{C}_k &=    (v_k^{-1}-\check{v}_k^{-1})\mathbf{R}_k
\end{IEEEeqnarray}
we then obtain that
\begin{IEEEeqnarray}{Cl}
           \calRk^{ub}  &\ge  a_k  -   (e_k^{(d)})^{-1} + ((v_k^{(d)})^{-1} -  (\check{v}_k^{(d)} )^{-1}) \sz
           \nonumber \\
           & +  (\mathbf{p}_k^{(d)})^H\mathbf{A}_k^{(d)} \mathbf{p}_k  +\mathbf{p}_k^H\mathbf{A}_k^{(d)}\mathbf{p}_k^{(d)}  - \mathbf{p}_k^H\mathbf{B}_k ^{(d)}\mathbf{p}_k \nonumber \\
           &  - \sumnok\Pl^H\mathbf{C}_k ^{(d)}\Pl  
\end{IEEEeqnarray} 
where the equality holds when we have that $\Pone,\Ptwo, \cdots, \PK=\PoneIterd,\PtwoIterd, \cdots, \PKIterd$.
For the objective function, we further have that
\begin{IEEEeqnarray}{Cl}
           f  &\ge  c_1^{(d)}
            + \sumK w_k(\mathbf{p}_k^{(d)})^H\mathbf{A}_k^{(d)} \mathbf{p}_k  +\sumK w_k\mathbf{p}_k^H\mathbf{A}_k^{(d)}\mathbf{p}_k^{(d)}  \nonumber \\
            &~~- \sumK w_k\mathbf{p}_k^H\mathbf{B}_k ^{(d)}\mathbf{p}_k  - \sumK w_k\sumnok\Pl^H\mathbf{C}_k ^{(d)}\Pl  
\end{IEEEeqnarray} 
where $c_1^{(d)}= \sumK(a_k  -   (e_k^{(d)})^{-1} + ((v_k^{(d)})^{-1} -  (\check{v}_k^{(d)} )^{-1}) \sz)$.
Let the matrix $\mathbf{D}_k$  be defined as
\begin{IEEEeqnarray}{Cl}  
          \mathbf{D}_k &=   w_k\mathbf{B}_k + \sumnok w_l\mathbf{C}_l.
           \label{eq:computation_of_Dk}
\end{IEEEeqnarray}
Let $g_1$ be a function defined as
\begin{IEEEeqnarray}{Cl}
           g_1 &=  c_1^{(d)}
            + \sumK w_k(\mathbf{p}_k^{(d)})^H\mathbf{A}_k^{(d)} \mathbf{p}_k  +\sumK w_k\mathbf{p}_k^H\mathbf{A}_k^{(d)}\mathbf{p}_k^{(d)}   \nonumber \\
            &~~~~- \sumK \mathbf{p}_k^H\mathbf{D}_k ^{(d)}\mathbf{p}_k.
\end{IEEEeqnarray} 
  Then, we obtain that
  \begin{IEEEeqnarray}{Cl}
           f \ge g_1
\end{IEEEeqnarray} 
  where the equality holds when $\Pone,\Ptwo, \cdots, \PK=\PoneIterd,\PtwoIterd, \cdots, \PKIterd$. Thus, $g_1$ is a minorizing function of $f$ at the points $\PoneIterd,\PtwoIterd, \cdots, \PKIterd$.

\section{Proof of Theorem \ref{th:minorizing_function_2}}
\label{sec:proof_of_minorizing_function_2} 
First, we can rewritten $g_1$ as
\begin{IEEEeqnarray}{Cl}
 g_1 &= c_1^{(d)}
            + \sumK w_k(\mathbf{p}_k^{(d)})^H\mathbf{A}_k^{(d)} \mathbf{p}_k  +\sumK w_k\mathbf{p}_k^H\mathbf{A}_k^{(d)}\mathbf{p}_k^{(d)}   \nonumber \\
            &- \sumK \mathbf{p}_k^H\mathbf{D} ^{(d)}\mathbf{p}_k 
            + \sumK \mathbf{p}_k^H(\mathbf{D}^{(d)}-\mathbf{D}_k^{(d)})\mathbf{p}_k.
            \label{eq:g1_new_form}
\end{IEEEeqnarray}
From $\mathbf{D} =    \sumK w_k\mathbf{C}_k$ and $\mathbf{D}_k =   w_k\mathbf{B}_k + \sumnok w_l\mathbf{C}_l$, we have that
 \begin{IEEEeqnarray}{Cl}
     \mathbf{D}^{(d)}-\mathbf{D}_k^{(d)} = w_k\mathbf{C}_k^{(d)} - w_k\mathbf{B}_k^{(d)}.
\end{IEEEeqnarray}
From  \eqref{eq:computation_of_Bk} and \eqref{eq:computation_of_Ck}, we then have
 \begin{IEEEeqnarray}{Cl}
     \mathbf{C}_k - \mathbf{B}_k &=(v_k^{-1}-\check{v}_k^{-1})\mathbf{R}_k
           -  v_k^{-1}\check{v}_k^{-1}\mathbf{R}_k\Pk\Pk^H\mathbf{R}_k \nonumber \\
           &=v_k^{-1}\check{v}_k^{-1}\Pk^H\mathbf{R}_k \Pk\mathbf{R}_k
           -  v_k^{-1}\check{v}_k^{-1}\mathbf{R}_k\Pk\Pk^H\mathbf{R}_k 
           \nonumber \\
           &= v_k^{-1}\check{v}_k^{-1}(\Pk^H\mathbf{R}_k \Pk \mathbf{I}- \mathbf{R}_k\Pk\Pk^H)\mathbf{R}_k.
\end{IEEEeqnarray}
Since $\mathbf{R}_k\Pk\Pk^H$ is an rank-1 matrix with eigenvalue
$\Pk^H\mathbf{R}_k\Pk$, we have $\Pk^H\mathbf{R}_k \Pk \mathbf{I}- \mathbf{R}_k\Pk\Pk^H$ is a positive semidefinite matrix. Thus, $\mathbf{C}_k - \mathbf{B}_k$ and $\mathbf{D}^{(d)}-\mathbf{D}_k^{(d)}$ are also positive semidefinite matrices.
Then, we obtain the item $\sum_{k=1}^K \mathbf{p}_k^H(\mathbf{D}^{(d)}-\mathbf{D}_k^{(d)})\mathbf{p}_k$ in 
 \eqref{eq:g1_new_form}  is a convex
quadratic function of the precoding vectors.
By using the property of convex functions, we obtain that
 \begin{IEEEeqnarray}{Cl}
 &\sumK \mathbf{p}_k^H(\mathbf{D}^{(d)}-\mathbf{D}_k^{(d)})\mathbf{p}_k
 \nonumber \\
 &~~\ge \sumK (\mathbf{p}_k^{(d)})^H(\mathbf{D}^{(d)}-\mathbf{D}_k^{(d)})\mathbf{p}_k^{(d)} 
 \nonumber \\
 &~~~~+ \sumK (\mathbf{p}_k^H-(\mathbf{p}_k^{(d)})^H)(\mathbf{D}^{(d)}-\mathbf{D}_k^{(d)})\mathbf{p}_k^{(d)}
 \nonumber \\
 &~~~~+ \sumK(\mathbf{p}_k^{(d)})^H(\mathbf{D}^{(d)}-\mathbf{D}_k^{(d)})(\mathbf{p}_k-\mathbf{p}_k^{(d)} ).
 \end{IEEEeqnarray}
where the equality holds when $\Pone,\Ptwo, \cdots, \PK=\PoneIterd,\PtwoIterd, \cdots, \PKIterd$.
To get further results, we compute $(\mathbf{D}^{(d)}-\mathbf{D}_k^{(d)})\mathbf{p}_k^{(d)}$ as
  \begin{IEEEeqnarray}{Cl}
 &(\mathbf{D}^{(d)}-\mathbf{D}_k^{(d)})\mathbf{p}_k^{(d)} \nonumber \\
  &= v_k^{-1}\check{v}_k^{-1}((\Pk^{(d)})^H\mathbf{R}_k \Pk^{(d)}\mathbf{I}- \mathbf{R}_k\Pk^{(d)}(\Pk^{(d)})^H)\mathbf{R}_k\Pk^{(d)} \nonumber \\
  &= 0.
  \end{IEEEeqnarray}
  Then, we obtain that
  \begin{IEEEeqnarray}{Cl}
 &\sumK \mathbf{p}_k^H(\mathbf{D}^{(d)}-\mathbf{D}_k^{(d)})\mathbf{p}_k
 \nonumber \\
 &~~~~ \ge \sumK (\mathbf{p}_k^{(d)})^H(\mathbf{D}^{(d)}-\mathbf{D}_k^{(d)})\mathbf{p}_k^{(d)}  
 \end{IEEEeqnarray}
 where the equality holds when $\Pone,\Ptwo, \cdots, \PK=\PoneIterd,\PtwoIterd, \cdots, \PKIterd$. Let $g_2$ be defined as
 \begin{IEEEeqnarray}{Cl}
 g_2 &= c_2^{(d)}
            + \sumK w_k(\mathbf{p}_k^{(d)})^H\mathbf{A}_k^{(d)} \mathbf{p}_k  +\sumK w_k\mathbf{p}_k^H\mathbf{A}_k^{(d)}\mathbf{p}_k^{(d)}  \nonumber \\
            &~~~~ - \sumK \mathbf{p}_k^H\mathbf{D} ^{(d)}\mathbf{p}_k  
\end{IEEEeqnarray}
where 
\begin{IEEEeqnarray}{Cl}
	c_2^{(d)} = c_1^{(d)} +  \sumK (\mathbf{p}_k^{(d)})^H(\mathbf{D}^{(d)}-\mathbf{D}_k^{(d)})\mathbf{p}_k^{(d)}.
\end{IEEEeqnarray}
We have $g_1 \ge g_2$, where the equality holds when $\Pone,\Ptwo, \cdots, \PK=\PoneIterd,\PtwoIterd, \cdots, \PKIterd$. Thus, $g_2$ is a minorizing function of $g_1$, and also is a minorzing function of $f$.

\bibliographystyle{IEEEtran}
\bibliography{IEEEabrv,this_reference}

\begin{IEEEbiography}
[{\includegraphics[width=1in,height=1.25in,clip,keepaspectratio]{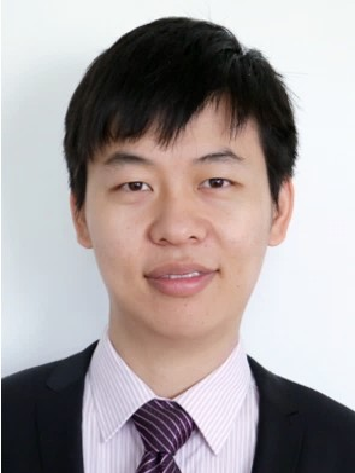}}]
{An-An Lu}(S'11-M'17)
received the B.E., M.E. and Ph.D. degree in electronic engineering
from Southeast University, Nanjing, China,
in 2006, 2012 and 2017, respectively.

Dr. Lu is a lecturer at the National
Mobile Communications Research Laboratory,
Southeast University, Nanjing, China.
From November 2014 to February 2016, he visited Missouri
University of Science and Technology, Rolla, MO, USA.
From 2006 to 2008,
he was with the Research Department, Hejian
Technology Co., Ltd., Suzhou, China.
His research interests include information theory and wireless communications.
\end{IEEEbiography}

\begin{IEEEbiography}[{\includegraphics[width=1in,height=1.25in,clip,keepaspectratio]{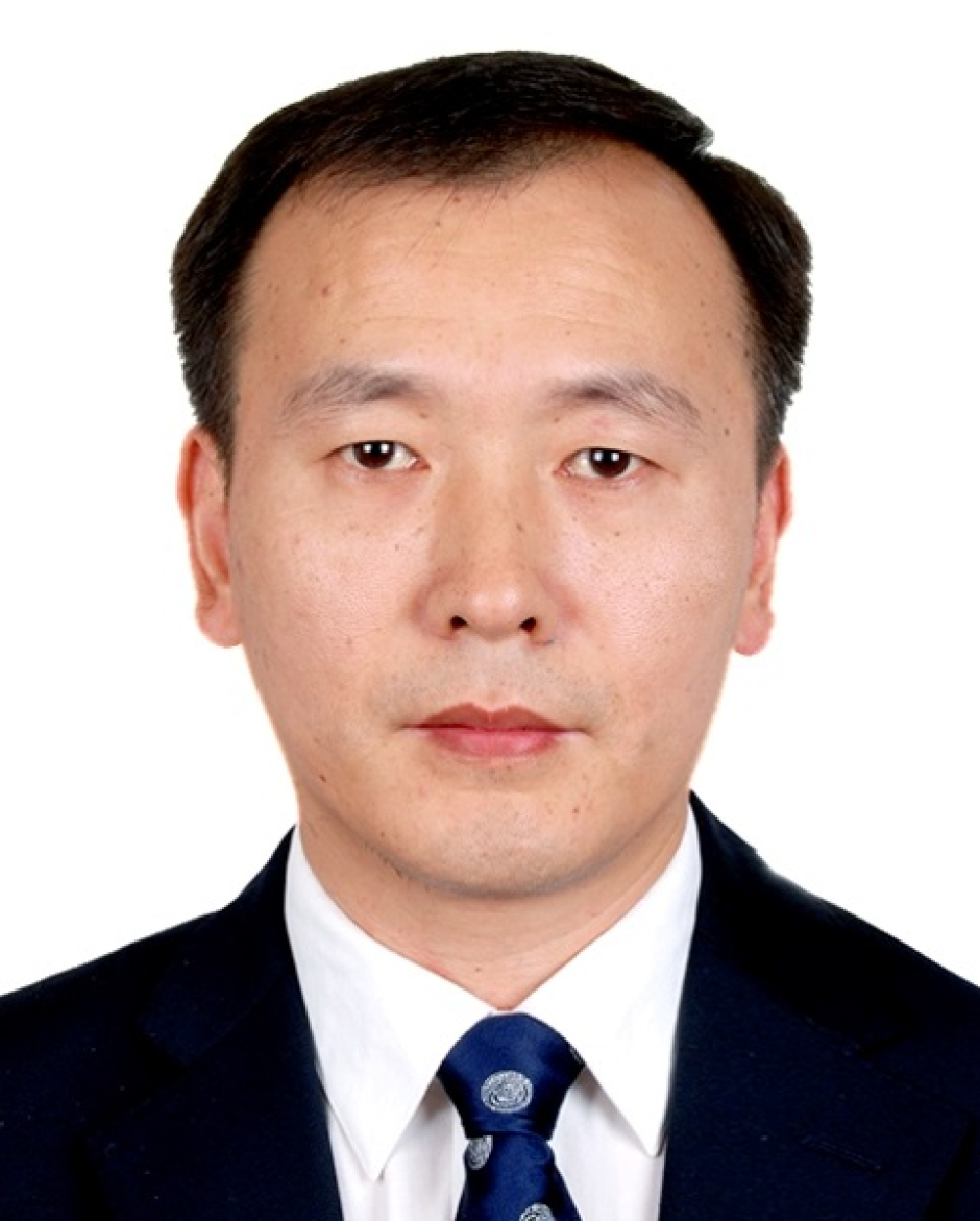}}]{Xiqi Gao}
(S'92--AM'96--M'02--SM'07--F'15) received the Ph.D. degree in electrical engineering from Southeast University, Nanjing, China, in 1997.

He joined the Department of Radio Engineering,
Southeast University, in April 1992. Since May 2001, he has been a professor of information systems and communications. From September 1999 to August 2000, he was
a visiting scholar at Massachusetts Institute of Technology, Cambridge, MA, USA, and Boston University, Boston, MA. From August 2007 to July 2008, he visited the Darmstadt
University of Technology, Darmstadt, Germany, as a Humboldt scholar. His current research interests include broadband multicarrier communications, massive MIMO wireless
communications, satellite communications, optical wireless communications, information theory and signal processing for wireless communications. From 2007 to 2012, he served as an Editor for the  IEEE Transactions on Wireless Communications. From 2009 to 2013, he served as an Editor for the  IEEE Transactions on Signal Processing. From 2015 to 2017, he served as an Editor for the  IEEE Transactions on Communications.

Dr. Gao was the recipient of  the Science and Technology Awards of the State Education Ministry of China in 1998, 2006 and 2009, the National Technological Invention Award of
China in 2011, the Science and Technology Award of Jiangsu
Province of China in 2014, and the 2011 IEEE Communications Society Stephen O. Rice Prize Paper Award in the Field of Communications Theory.
\end{IEEEbiography}

\begin{IEEEbiography}
[{\includegraphics[width=1in,height=1.25in,clip,keepaspectratio]{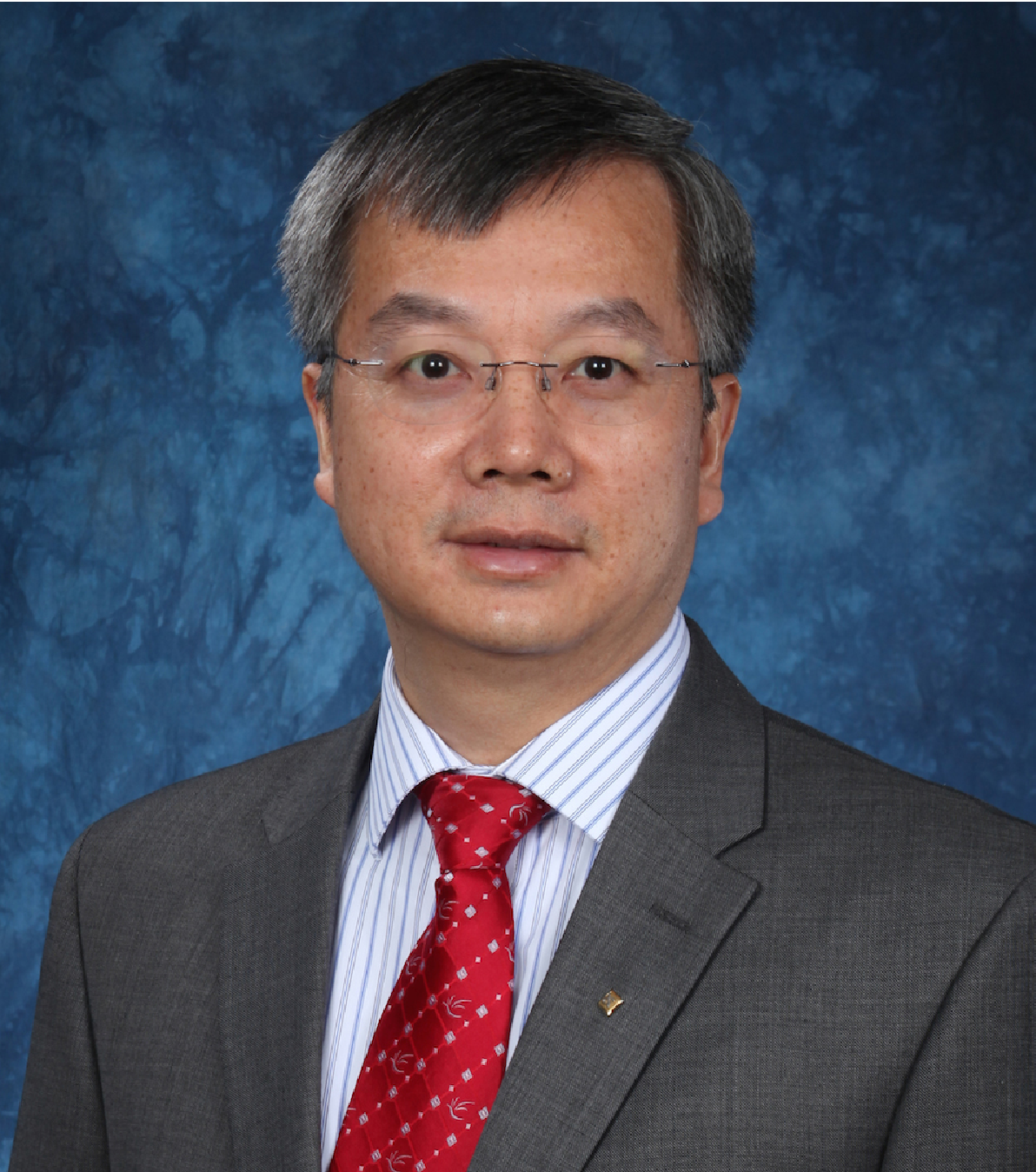}}]
{Chengshan Xiao}
(M'99--SM'02--F'10) received a Bachelor of Science degree from University of Electronic Science and Technology of China, a Master of Science degree from Tsinghua University, and a Ph.D. in electrical engineering from University of Sydney.

Dr. Xiao is the Chandler Weaver Professor and Chair of Electrical and Computer Engineering Department at Lehigh University. He is a Fellow of the IEEE and a Fellow of the Canadian Academy of Engineering. Previously, he served as a Program Director with the Division of Electrical, Communications and Cyber Systems at the USA National Science Foundation. He was a senior member of scientific staff with Nortel Networks, Ottawa, Canada, a faculty member at Tsinghua University, Beijing, China, University of Alberta, Edmonton, Canada, the University of Missouri - Columbia, MO, and Missouri University of Science and Technology, Rolla, MO. He also held visiting professor positions in Germany and Hong Kong. His research interests include wireless communications, signal processing, and underwater acoustic communications. He is the holder of several patents granted in USA, Canada, China and Europe. His invented algorithms have been implemented into Nortel's base station radio products after successful technical field trials and network integration.

Dr. Xiao is the Vice President for Publications of the IEEE Communications Society. Previously, he was the Awards Committee Chair, an elected Member-at-Large of Board of Governors, a member of Fellow Evaluation Committee, Director of Conference Publications, Distinguished Lecturer of the IEEE Communications Society. He also served as an Editor, Area Editor and the Editor-in-Chief of the IEEE Transactions on Wireless Communications, an Associate Editor of the IEEE Transactions on Vehicular Technology, and of the IEEE Transactions on Circuits and Systems-I. He was the Technical Program Chair of the 2010 IEEE International Conference on Communications, Cape Town, South Africa, a Technical Program Co-Chair of the 2017 IEEE Global Communications Conference, Singapore. He served as the founding Chair of the IEEE Wireless Communications Technical Committee. He received several distinguished awards including 2014 Humboldt Research Award, 2014 IEEE Communications Society Joseph LoCicero Award, 2015 IEEE Wireless Communications Technical Committee Recognition Award, and 2017 IEEE Communications Society Harold Sobol Award. 
\end{IEEEbiography}

\end{document}